\documentclass[11pt,aps,amsmath,amssymb,nofootinbib,notitlepage,longbibliography]{revtex4-1}
\usepackage{amsmath,amssymb}
\usepackage{slashed}
\usepackage{graphicx}
\usepackage{bm}
\usepackage[top=1.0in,bottom=1.0in,left=1.0in,right=1.0in]{geometry}
\usepackage[colorlinks,linkcolor=blue,citecolor=blue]{hyperref}

\newcommand{\be}{\begin{equation}}
\newcommand{\ee}{\end{equation}}
\newcommand{\bea}{\begin{eqnarray}}
\newcommand{\eea}{\end{eqnarray}}
\newcommand{\ba}{\begin{eqnarray}}
\newcommand{\ea}{\end{eqnarray}}

\newcommand{\Dslash}{D\hspace{-1.6ex}/\hspace{0.6ex} }

\begin{document}

\title{
Small size instanton contributions to the\\
quark quasi-PDF and matching kernel}

\author{Yizhuang Liu}
\email{yizhuang.liu@sjtu.edu.cn}
\affiliation{Tsung-Dao Lee Institute, Shanghai Jiao Tong University, Shanghai, 200240, China\\
 Institut  fur Theoretische Physik, Universitat Regensburg, D-93040 Regensburg, Germany}

\author{Ismail Zahed}
\email{ismail.zahed@stonybrook.edu}
\affiliation{Center for Nuclear Theory, Department of Physics and Astronomy, Stony Brook University, Stony Brook, New York 11794--3800, USA}



\begin{abstract}
We investigate the non-perturbative contribution of instantons  to current matching kernels
used in the context of the large momentum effective theory (LaMET). We derive explicitly these contributions
using first principle semi-classical calculus for the unpolarized and polarized quark parton distributions and the matching kernel,
and show that they are part of a trans-series expansion. These contributions  are substantial  at current lattice matching momenta.
\end{abstract}

\maketitle

\section{Introduction}

Light cone distributions are central to the description of hard inclusive and exclusive processes. Thanks to factorization,
a hard process factors  into a perturbatively calculable contribution times pertinent parton distribution and fragmentation
functions. Standard examples can be found in deep inelastic scattering, Drell-Yan process and jet production to cite a few.
The parton distribution functions are defined on the light front, and their moments usually fitted using large empirical data bank.  They
are not readily amenable  to a non-perturbative and first principle formulation using lattice simulations.

 Ji~\cite{Ji:2013dva} has put forth the concept of quasi-parton distributions as a way to approach the light cone parton distributions
from first principles on the lattice~\cite{Zhang:2017bzy,Ji:2015qla,Bali:2018spj,Alexandrou:2018eet,Izubuchi:2019lyk,Izubuchi:2018srq}.
The idea is to start from pertinent partonic correlators at equal-time in a boosted hadron state, which
are amenable to an  Euclidean formulation. Under increasing boosts, these quasi-distributions asymptote the light cone distributions
space-like. They are conjectured to map onto their time-like analogue through perturbative matching.  This conjecture can be checked
to hold non-perturbatively in two-dimensional QCD and at sub-leading order in the large number of color limit~\cite{Ji:2018waw}. Some  variants of this
formulation can be found in the form of pseudo distributions~\cite{Radyushkin:2017gjd}, and lattice cross sections~\cite{Ma:2014jla}.

A  number of QCD lattice collaborations have implemented some of these ideas with some reasonable success in extracting
the light cone parton distributions through matching. Unfortunately, most simulations become increasingly noisy as the boosting
is increased, making the extraction by matching limited to boosting momenta of the order of few GeV. In this range, the power
corrections and non-perturbative effects are still sizable even in the matching kernel. The chief motivation of this analysis is
to estimate some of these non-perturbative contributions from first principles, using QCD semi-classics.

Several QCD lattice simulations have shown  that the bulk characteristics and correlations  in the QCD vacuum
are mostly unaffected by lattice cooling~\cite{Chu:1994vi} where quantum effects are pruned,  suggesting that semi-classical gauge
and fermionic fields dominate  the ground state structure. At weak coupling, instantons and anti-instantons
are exact classical gauge tunneling configurations with large actions and finite topological charge which
support exact quark zero modes with specific chirality.

The large size instanton and anti-instanton configurations in the QCD vacuum are at the origin of the spontaneous breaking
of chiral symmetry and the origin of mass. They still lack the long-range characteristics of the gauge fields necessary for the
disordering of the large Wilson loops in the quenched approximation, although some form of screening in the full theory may
not require that.

The small size instantons and anti-instantons configurations contribute ideally to short distance processes through weak coupling
semi-classics, although they become increasingly rare as their density wanes out.  It is the purpose of this work to explore their
contribution to matching kernels in the quasi-parton approach~\cite{Ji:2020ect}, following  the
recently suggested program to consider the role of small size instantons in semi-inclusive processes~\cite{Shuryak:2020ktq}.
Our analysis readily extends to the pseudo-parton program and lattice cross sections.

The outline of the paper is as follows: in section~\ref{sec_instantons} we briefly review some bulk aspects of the large and small
instantons that are relevant for this work, with a particular discussion on the size averaging procedure. In section~\ref{section_LSZ}
we provide results for two complementary LSZ reductions in Euclidean space. In section~\ref{sec_unpolarized} we derive the
contribution of  a random ensemble of instantons and anti-instantons to the unpolarized quark parton distributions.
In section~\ref{sec_polarized} the results are extended to the polarized distributions where the mixing between zero
modes and non-zero modes is dominant. In section~\ref{sec_loop} a one-loop gluon effect in the instanton background
is considered and shown to yield a large correction to the perturbative matching kernel.  Our
conclusions are in section~\ref{sec_conclusions}.

\section{Instanton effects}
\label{sec_instantons}

The bulk characteristics of the instanton liquid model are captured by the parameters~\cite{Shuryak:1981ff}

\begin{equation}
n_{I+\bar I}\approx 1/L^4\approx 1/{\rm fm}^{4} \qquad\qquad \rho\approx (1/3) \, {\rm fm} \approx  1/(0.6\, {\rm GeV})  \label{eqn_ILM}
\end{equation}
for the instanton plus anti-instanton density and size, respectively. They combine in the  dimensionless packing parameter
$\kappa\equiv \pi^2\rho^4 n_{I+\bar I}\approx 0.1$,
a measure of  the diluteness of the instanton-anti-instanton ensemble in the QCD vacuum. Previous
lattice simulations using cooling methods support these observations, see \cite{Schafer:1996wv} for a review.

A more recent lattice simulation aimed at extracting the running gauge coupling constant with momentum~\cite{Athenodorou:2018jwu},
through a pertinent  ratio of 2- and 3-point gluonic functions, has suggested a larger instanton density with
$\kappa\approx 1$. This observation at large running momenta, points at the effects from the smaller
size instanton-anti-instanton configurations which are likely to affect short distance physics such as the running
of the coupling, with negligible effects on the longer range mechanism of chiral symmetry breaking.

Small size instanton fields are strong, since their field strength  is large.
In fact, even for dominant and large  size instantons  with $\rho\approx 0.30\, {\rm fm}$ typical for chiral symmetry breaking,
it is still sizable, with typicall chromo-electric and -magnetic fields $E=B\approx \sqrt{48}/\rho^2\approx 2.5\,{\rm GeV}^2$.
These fields are
%
 comparable to the current momentum extrapolations using LaMET in the hard matching kernel. Their
short distance contribution can be assessed using semi-classics.
For many estimates it is sufficient to use  the fixed  value of $\rho\approx 0.30\, {\rm fm}$. However,  in harder kernels
the small size instantons are dominant, but their density is suppressed~\cite{Hasenfratz:1999ng,Shuryak:1999fe}

\begin{equation}
\label{dn_dist}
\frac{dn(\rho)}{d\rho} \approx   {1 \over \rho^5}\big(\rho \Lambda_{QCD} \big)^{b_{QCD}} \, e^{- C\rho^2/L^2}
\end{equation}
with $b_{QCD}=11N_c/3-2N_f/3\approx 9$ (one loop) and  $C$ a number of order 1.
 This notwithsanding, in this regime the effective gauge coupling is weak, the action is
large and the use of the semi-classical approximation is justified, with the instanton contribution leading and the gluon
exchange contribution subleading. We will limit the hard instanton contributions to $\rho\leq \rho_S$ in the hard kernel, and relegate
the contributions from larger instantons with $\rho\geq \rho_S$ to the wavefunctions~\cite{Kock:2020frx}.

\section{LSZ reduction in instanton background}\label{section_LSZ}

We start by recalling briefly  the LSZ reduction formula for the instanton. We note that the LSZ reduction is not guaranteed unless the Euclidean field theory has an Hamiltonian interpretation, which is lacking in the instanton model of the QCD vacuum. Here it is understood as an algorithm, following the intial suggestions in~\cite{Balitsky:1993ki,Moch:1996bs}. We will present the reduction in two different limits: 1/ the zero Euclidean momentum limit; 2/ the large Euclidean time limit. Both construction yield similar results modulo an infrared sensitivity noted only in the first approach. The first reduction was recently used in the discussion of the 
instanton contributions to the  mesonic form-factors~\cite{Shuryak:2020ktq}.

\subsection{Fermion propagator in an instanton}

The full non-zero propagator in the instanton background in the chiral-split form reads~\cite{Brown:1977eb}

\begin{eqnarray}
\label{BROWN}
S_{nz}(x,y)= &&\overrightarrow{\Dslash_x}\Delta(x,y) {1+\gamma_5 \over 2} +  \Delta(x,y) \overleftarrow{ \Dslash}_y {1-\gamma_5 \over 2} \nonumber\\
=&& \overline{S}(x,y) \frac{1+\gamma_5}2+{S}(x,y)\frac{1-\gamma^5}2
\end{eqnarray}
with the free Weyl  propagators $S_0=1/\overline{\partial}$ and $\overline{S}_0=1/{\partial}$, in the notations detailed in Appendix~\ref{app_defs}.
The long derivative $\slashed{D}=\slashed{\partial}-i\slashed{A}$ acts on the left and right respectively of the (massless) scalar
propagator,

\begin{equation}
\label{PROSCALAR}
\Delta(x,y) =\Delta_0(x-y)\bigg(1+ \rho^2 \frac{[Ux \bar y U^\dagger]}{x^2 y^2}\bigg) \frac 1{(\Pi_x \Pi_y)^{\frac 12}}
\end{equation}
with $\Delta_0(x)=1/(2\pi x)^2$ the free scalar propagator, and $\Pi_x=1+\rho^2/x^2$, and
$x,\bar y$ are convoluted with (Euclidean 4d) sigma matrices (\ref{eqn_4dsigma}).
Each explicit contribution is

\begin{eqnarray}
\label{BROWNCHIRAL}
\overline{S}(x,y)=\bigg(\overline{S}_0(x-y)\bigg(1+\rho^2\frac {[Ux\bar yU^\dagger]}{x^2y^2}\bigg)+\frac{\rho^2\overline{\sigma}_\mu}{4\pi^2}
\frac{[Ux\overline{\sigma}_\mu( x- y)\bar yU^\dagger]}{\Pi_xx^4(x-y)^2y^2}\bigg)\,\frac 1{(\Pi_x\Pi_y)^{\frac 12}} \nonumber\\
S(x,y)=\bigg(S_0(x-y)\bigg(1+\rho^2\frac {[Ux\bar yU^\dagger]}{x^2y^2}\bigg)+\frac{\rho^2\sigma_\mu}{4\pi^2}
\frac{[Ux(\bar x-\bar y)\sigma_\mu\bar yU^\dagger]}{x^2(x-y)^2y^4\Pi_y}\bigg)\,\frac 1{(\Pi_x\Pi_y)^{\frac 12}}
\end{eqnarray}
and with $U$ valued in $SU(N_c)$.
When a mixture of color and spinor indices occurs, the spinor matrices act on the upper left corner of the
$N_c\times N_c$ color matrices.  The zero modes and their LSZ reduction are discussed in Appendix~\ref{app_zero}.

The effects of the quark masses on the quark propagator in the instanton or anti-instanton fields, are  not known in closed form. However, for small masses, the propagator
can be expanded around the chiral limit, and in the instanton field it reads

\be
\label{MASS}
\frac{\Psi_0(x,)\Psi_0^\dagger(y)}{im}+S_{nz}(x,y)-im\bigg( \Delta(x,y){1+\gamma_5 \over 2}+\int d^4z\overline S(x,z)S(z,y){1-\gamma_5 \over 2}\bigg)+{\cal O}(m^2)
\ee
Note the chiral mixing induced by the mass $m$.
In the expansion of correlators, the ${\cal O}(m)$ term may combine with the ${\cal O}(1/m)$  term, to yield a
finite ${\cal O}(m^0)$ contribution. This will be the case below for the unpolarized quark quasi-PDF.

\subsection{Euclidean zero momentum limit}

The LSZ reduction of the quark propagator $S(x,y)$ in an instanton background, is obtained by
say taking its half  Fourier transform $S(k,y)$, reducing it through $\slashed S(k,y)$, expanding
the result around the Euclidean point $k^2\approx 0$  and analytically continuing the result to Minkowski
space. In the chiral split form (\ref{BROWNCHIRAL}),
the reduction for  $S\bar k$ and $k\overline S$ give~\cite{Moch:1996bs}

\begin{eqnarray}
\label{ZM-ZM-3}
ik\overline{S}(k,z_+)&\approx& \frac{e^{+ik\cdot z_+}}{\Pi^{1/2}_+}
\bigg(1+\frac {\rho^2}{2z_+^2}\frac{Uk\bar{z}_+U^\dagger}{k\cdot z_+}(1-e^{-ik\cdot z_+})\bigg)
\nonumber\\
{S}(z_-,k)i\bar k&\approx& \frac{e^{-ik\cdot z_-}}{\Pi^{1/2}_-}
\bigg(1+\frac {\rho^2}{2z_-^2}\frac{Uz_-\bar{k}U^\dagger}{k\cdot z_-}(1-e^{+ik\cdot z_-})\bigg)
\end{eqnarray}
with $z_\pm =(\pm z/2, z_\perp)$, while for  the more involved reductions
$\overline S k$ and $\bar k S$ one has~\cite{Shuryak:2020ktq}

\begin{eqnarray}
\label{ZM-ZM-3X}
\overline{S}(z_-,k)ik&\approx&
 \frac{e^{-ik\cdot z_-}}{\Pi^{1/2}_-}
 \bigg(1-\frac{\rho^2}{2z_-^2}\frac{U k\bar z_- U^\dagger}{k\cdot z_-}\bigg(1-\frac {i}{k\cdot z_-}(1-e^{+ik\cdot z_-})\bigg)
- \frac {\rho^2}{m^2}\frac {ik}{z_-^4\Pi_-}\bar\sigma_\mu U {z}_- \bar\sigma_\mu U^\dagger\bigg)
\nonumber\\
i\bar k{S}(k,z_+)&\approx&
\frac{e^{+ik\cdot z_+}}{\Pi^{1/2}_+}
 \bigg(1-\frac{\rho^2}{2z_+^2}\frac{Uz_-\bar k U^\dagger}{k\cdot z_+}\bigg(1+\frac i{k\cdot z_+}(1-e^{-ik\cdot z_+})\bigg)
- \frac {\rho^2}{m^2}\frac {i\bar k}{z_+^4\Pi_+}\sigma_\mu U \sigma_\mu  \bar{z}_+U^\dagger\bigg)\nonumber\\
\end{eqnarray}
Note that the  Euclidean regulator $-k^2\approx 0\rightarrow m^2$ was used in the last contributions
appearing in (\ref{ZM-ZM-3X}). This infrared sensitivity will be circumvented below through an alternative reduction
scheme that is more commensurate with QCD lattice formulations.

\subsection{Euclidean large   time limit}

An alternative reduction scheme that will prove to be infrared safe with an almost identical outcome, consists in taking the
large Euclidean  time asymptotics instead, to put the quark on mass shell, a common procedure on the lattice. Specifically,  the right reduction of $\bar S$  is

\be
\label{T0}
\lim_{T\rightarrow - \infty}\int d^3\vec{y} e^{-i\vec{p}\cdot \vec{y}}\bar S(x;\vec{y},T)
\ee
to bring the massless quark on the energy shell without recourse to the Fourier transform and the external leg reduction.
 With this in mind, and collecting the results (\ref{T2}-\ref{T4}) from Appendix~\ref{app_large},
 we obtain the large time LSZ reduced results for the non-zero mode  $\bar S$
of the quark propagator in an instanton background

\begin{align}
\label{T6}
&\int d^3\vec{y} e^{-i\vec{p}\cdot \vec{y}}\bar S(x;\vec{y},T)=\frac{e^{-|T||\vec{p}|}e^{-T_1|\vec{p}|-i\vec{p}\cdot \vec{x}}}{\Pi_x^{\frac{1}{2}}}\nonumber \\
&\times \bigg(\frac{1}{2}{\cal P}_--\frac{1}{4x^2}{\cal P}_-|\vec{p}|\rho^2Ux{\cal P}_-U^{\dagger}I_-^1-\frac{\rho^2}{4x^4\Pi_x}\bar \sigma_{\mu}Ux\bar\sigma^{\mu}x{\cal P}_-U^{\dagger}I_--\frac{\rho^2}{2x^4|\vec{p}|\Pi_x}\bar \sigma_{\mu}Ux\bar\sigma^{\mu}U^{\dagger}\bigg) \ ,
\end{align}
with

\begin{align}
&I_-^n=\int_0^1 dt t^ne^{i(1-t)\vec{x}\cdot \vec{p}+(1-t)T_1|\vec{p}|}\nonumber\\
&I_-=\int_0^1 dt e^{i(1-t)\vec{x}\cdot \vec{p}+(1-t)T_1|\vec{p}|}
\end{align}
Similarly, we have

\begin{align}
\label{T7}
&\int d^3\vec{y}\bar S(T,\vec{y};x)e^{i\vec{p}\cdot \vec{y}}=\frac{e^{-|T||\vec{p}|}e^{T_1|\vec{p}|+i\vec{p}\cdot \vec{x}}}{\Pi_x^{\frac{1}{2}}}\nonumber \\
&\times \bigg(\frac{1}{2}{\cal P}_-+\frac{1}{4x^2}{\cal P}_-|\vec{p}|\rho^2U{\cal P}_+\bar xU^{\dagger}I_+^1+\frac{\rho^2}{8x^2}|\vec{p}|\bar \sigma_{\mu}U{\cal P}_{+}\bar \sigma^{\mu}{\cal P}_{+}\bar xU^{\dagger}I_+\bigg) \ ,
\end{align}
with

\begin{align}
&I_+^n=\int_{0}^1dt\, t^n e^{-i(1-t)\vec{x}\cdot \vec{p}-(1-t)T_1|\vec{p}|}\nonumber\\
&I_+=\int_{0}^1dt\,(1-t)e^{-i(1-t)\vec{x}\cdot \vec{p}-(1-t)T_1|\vec{p}|}
\end{align}
The helicity
projectors are
${\cal P}_{\pm}=1\pm\frac{\vec{\sigma}\cdot \vec{p}}{|\vec{p}|}$. The results for the anti-instanton follow by conjugation. Note that the reduction (\ref{T6})
is infrared safe in contrast to (\ref{ZM-ZM-3X}). We will use it for the analysis to follow.

For completeness, the large time asymptotics for the scalar propagator in the instanton background in the mass expansion (\ref{MASS})

\be
\label{MASS1}
\lim_{T\rightarrow - \infty}\int d^3\vec{y} e^{-i\vec{p}\cdot \vec{y}} \Delta (x;\vec{y},T)
\ee
follows from the same reasoning with the result

\begin{align}
\label{MASS2}
&\int d^3\vec{y} e^{-i\vec{p}\cdot \vec{y}}\Delta (x;\vec{y},T)=\frac{e^{-|T||\vec{p}|}e^{-T_1|\vec{p}|-i\vec{p}\cdot \vec{x}}}{\Pi_x^{\frac{1}{2}}}\,
\bigg(\frac {2\pi^2}{|\vec p|}-\frac{\pi^2\rho^2}{x^2}{\cal P}_-UxU^\dagger I_+^0\bigg)
\end{align}
and similarly for $ \Delta (T,\vec{y}; x) $ with the exchange $x\rightarrow \bar x$ in the last contribution.

 \begin{figure}[h!]
	\begin{center}
		\includegraphics[width=16cm]{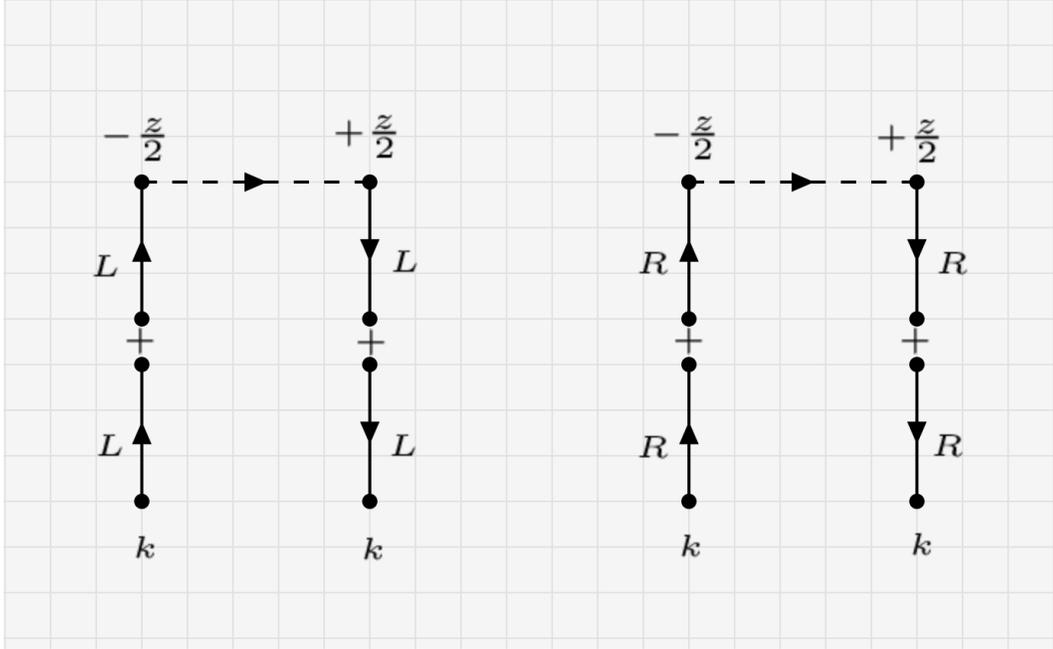}
		\caption{Non-vanishing small instanton contributions to twist-2 matching factor from only the non-zero modes.
		}
		\label{fig-NZM}
	\end{center}
\end{figure}

\section{Instanton contribution to the quasi-PDF: tree level}
\label{sec_unpolarized}

Given the diluteness of the QCD instanton vacuum with a small packing fraction $\kappa<1$, it will be sufficient to assess the small size instanton and anti-instanton
contributions to the leading twist unpolarized and polarized partonic distributions using the single instanton approximation (SIA)~\cite{Andrei:1978xg,Faccioli:2001ug}.
The approximation is limited to distances which are smaller than the mean distance $L\approx 1$ fm in (\ref{eqn_ILM}),
the distance to the nearest neighbor instanton or anti-instanton in the QCD instanton vacuum.
This will be understood throughout.
We will carry first this analysis at tree level and correct it to one-loop. The latter correction is important for considerations in relation to the matching kernel in LaMET
beyond perturbation theory.

\subsection{Tree Contribution to the unpolarized quark-PDF}~\label{unpolarized}

The analysis of the quasi-PDF of the light mesons in the context of the random instanton vacuum using
the planar approximation,  was discussed recently in~\cite{Kock:2020frx}.
Here  we study the SIA contribution to the unpolarized quark-PDF. The
instanton  non-zero mode contributions are illustrated in Fig.~\ref{fig-NZM} and read

\begin{align}\label{nZM1}
&\frac{n_{I+\bar I}}{2}\int d^4z_{I} \overline \chi_R(P)i\bar P{S}(P,\frac{z}{2}-z_I)\bar \sigma^z[+\frac{z}{2},-\frac{z}{2}]_I{S}(-\frac{z}{2}-z_I,P)i\bar P\chi_R(P)\nonumber \\
+&\frac{n_{I+\bar I}}{2}\int d^4z_{I}\overline\chi_L(P)i P{S}(P,\frac{z}{2}-z_I)\sigma^z
[+\frac{z}{2},-\frac{z}{2}]_{I}{S}(-\frac{z}{2}-z_I,P)i P\chi_L(P) \ ,
\end{align}
The anti-instanton contributions follow similarly.
The PDF follows from the double limit of
$z^2\rightarrow 0$ and $P^z\rightarrow\infty$ but fixed $zP^z$.  Hence, $z\approx 1/P^z\ll L$
justifying the use of the SIA  with  small size instantons.
The gauge link  can be reduced by noting that it Abelianizes along $\sigma_{z\nu}z_{I\nu}$,

\be
\label{U1}
[+\frac z2, -\frac z2]_I={\bf P}\,e^{-i\int_{-\frac z2-z_{3I}}^{+\frac z2-z_{3I}}A^U_3(x_3-\tilde z_I)dx_3}
={\rm cos}F(z, z_I)+i\,U^\dagger 2\sigma_{z\nu}\hat z_{I\nu}U\,{\rm sin}F(z, z_I)
\ee
with  $\bar \eta^a_{\mu\nu}T^a=\frac 1{4i}(\bar \sigma_\mu\sigma_\nu-\bar\sigma_\nu\sigma_\mu)=\sigma_{\mu\nu}$,
$\tilde z_I= (z_{\perp I},0,z_{4I})\equiv (z_{1I}, z_{2,I}, 0,z_{4I})$, and

\begin{align}
\label{U2}
F(z,z_I)=&\frac 12\bigg[\frac {\sqrt{ z^2_{\perp I}}}{\sqrt{\tilde z_{I}^2}}\,\bigg(
{\rm tan}^{-1}\bigg(\frac {z/2-z_{3I}}{\sqrt{\tilde z_{I}^2}}\bigg)+{\rm tan}^{-1}\bigg(\frac {z/2+z_{3I}}{\sqrt{\tilde z_{I}^2}}\bigg)\bigg)\nonumber\\
&-
\frac {\sqrt{z^2_{\perp I}}}{\sqrt{\tilde z_{I}^2+\rho^2}}\bigg(
{\rm tan}^{-1}\bigg(\frac {z/2-z_{3I}}{\sqrt{\tilde z_{ I}^2+\rho^2}}\bigg)+{\rm tan}^{-1}\bigg(\frac {z/2+z_{3I}}{\sqrt{\tilde z_{ I}^2+\rho^2}}\bigg)\bigg)
\bigg]
\end{align}
To project onto the left- or right-handed Dirac spinors in (\ref{nZM1}), it is sufficient to keep only the terms proportional to ${\cal P}_{\pm}$
 in the LSZ reduced quark propagators (\ref{T6}-\ref{T7}), with the result

\begin{align}
\label{U3}
\tilde f(z,P^z)= 2\times e^{izP^z}{\rm Tr}_c\bigg[\frac{2}{{\sqrt{\Pi_{+}}\sqrt{\Pi_{-}}}}\left(A_+\tilde A_-\right) \bigg]
\end{align}
Note that the overall factor of 2 takes care of the anti-instanton contribution, with the definitions (before subtraction)

\begin{align}
\label{U4}
A_+=&\frac{1}{2}+\frac{\rho^2}{4z_+^2}U\bar p \bar z_+ U^{\dagger}I_+^1 \ , \nonumber\\
\tilde A_-=&\frac{1}{2}+\frac{\rho^2}{4z_-^2}Uz_- pU^{\dagger}I_-^1
\end{align}
and ($z_\pm=\pm z/2-z_I$)

\be
\Pi_\pm=\frac 1{1+\frac{\rho^2}{(z_\pm-z_I)^2}}=\frac 1{1+\frac{\rho^2}{(\pm z/2-z_I)^2}}
\ee
Using the results  ${\rm Tr}_c(\bar p \bar z)={\rm Tr}_cP^z(1-\sigma^z)(i\sigma^z z)=-2izP^z$, ${\rm Tr}_c(p z)=2iP^zz$ and $p\bar p=p^2=0$,
we can simplify (\ref{U3})

\begin{align}
{\rm Tr}_c A_+\tilde A_-=\frac{1}{2}-\frac{\rho^2}{2N_cz_+^2}(iP^zz_{+z})(I_+^1) +\frac{\rho^2}{2N_cz_-^2}(iP^zz_{-z})(I_-^1)
\end{align}
and obtain (after subtraction)


\begin{align}
\label{qpdf1}
\tilde f(z^2,zP^z)=&\,{n_{I+\bar I}}\int d^4z_I\bigg[ e^{-izP^z}\bigg(\frac 1{\sqrt{\Pi_+}}\frac 1{\sqrt{\Pi_-}}-1\bigg)\,{\rm cos}F(z, z_I)\nonumber\\
 &\qquad\qquad-\frac 1{2N_c}\bigg(\frac{\rho^2}{z_+^2}\frac{(iP^zz_{+})}{\sqrt{\Pi_+\Pi_-} } -\frac{\rho^2}{z_-^2}\frac{(iP^zz_{-})}{\sqrt{\Pi_+\Pi_-} }\bigg){\rm cos}F(z,z_I)\,
\int_{0}^1  dt\,te^{-i(t+1)zP^z/2}\bigg] \nonumber\\
\end{align}
The integration over the $z_I$-position diverges quadratically,  following  the slow decay of the 2-point function in a quark on mass-shell.
As noted earlier, this divergence is  cutoff by the mean distance $L$ to the nearest neighbor instanton or anti-instanton in the SIA, a simple
way to factor in the effects of the inter-instanton interactions.
With this in mind and since $F\rightarrow 0$ as $z\rightarrow 0$, the result is

\be
{{n_{I+\bar I}}}\int d^4z_I \bigg(\frac 1{\sqrt{\Pi_+}}\frac 1{\sqrt{\Pi_-}}-1\bigg){\rm cos}{F(z,z_I)}\approx -\frac{\pi\sqrt\kappa}2
+{\cal O}(z^2)
\ee
where we used that  $n_{I+\bar I}=1/L^4$ for  a diluteness parameter $\kappa=\pi^2\rho^4n_{I+\bar I}$.
The transmutation of the expansion from $\kappa\rightarrow \sqrt{\kappa}$ reflects on the the screening-like
effect,  and is analogous to the one noted in~\cite{Kock:2020frx}. Similarly we have

\be
\frac{n_{I+\bar I}}{N_c}\int d^4z_I \frac{\rho^2}{z_\pm^2}\frac {P^zz_{\pm }}{\sqrt{\Pi_+\Pi_-}}\,{\rm cos}{F(z,z_I)}\approx
\pm \frac {zP^z}{2N_c} \,{n_{I+\bar I}}\,\int d^4z_I \frac{\rho^2}{z_\pm^2}\frac 1{\sqrt{\Pi_+\Pi_-}}
\approx   \pm \frac {zP^z}{2N_c} \pi\sqrt{\kappa}+{\cal O}(z^2)
\ee
In terms of  the integral  transform

\begin{align}
\tilde f(z^2,\lambda=zP^z)=\int_{0}^1 dx e^{-ix \lambda }\tilde f(z^2,x)
\end{align}
the final one-loop instanton result at tree level is

\begin{align}
\tilde f(z^2,x)\approx\frac{\pi\sqrt\kappa}2\bigg(
\theta(x-1/2)\theta(1-x) -\delta(1-x)\bigg)+{\cal O}(z^2, 1/N_c, \kappa)\rightarrow  f(x)
\end{align}
This has the correct support in $[0,1]$,  and identifies with the light-cone
PDF $f(x)$ as $z^2\rightarrow 0$. The additional contributions to (\ref{qpdf1}) stemming from
the remaining cross terms in (\ref{nZM1}) with a similar behavior, are listed in Appendix~\ref{ALL} both
for the $z=0$ and the leading $z\neq 0$ terms for completeness.

\subsection{Mixing}

There is an additional mixed zero-mode and non-zero-mode contribution
following from the cross contribution from the  ${\cal O}(m)$ and ${\cal O}(1/m)$ in (\ref{MASS}), with the result
to order ${\cal O}(m^0)$

\begin{align}
\label{ZMZ2}
&\frac{n_{I+\bar I}}{2}\int d^4z_{I}\int dU \,e^{iP^zz}\bigg(\frac{4\pi^2\rho^3}{\Pi_+^{\frac 32}\Pi_-^{\frac 12}}\bigg)
\bigg[(\overline\chi_R(P)\epsilon U)(U^\dagger \epsilon \bar S_0(z_+))\sigma^z\nonumber\\
&\times
({\rm cos}F(z, z_I)+i\,U^\dagger 2\sigma_{z\nu}\hat z_{I\nu}U\,{\rm sin}F(z, z_I))\bigg(\frac {2\pi^2}{|\vec p|}-\frac{\pi^2\rho^2}{z_-^2}{\cal P}_-Uz_-U^\dagger I_+^0\bigg)\chi_R(P)\bigg]\nonumber\\
+&\frac{n_{I+\bar I}}{2}\int d^4z_{I}\int dU \,e^{iP^zz}\bigg(\frac{4\pi^2\rho^3}{\Pi_-^{\frac 32}\Pi_+^{\frac 12}}\bigg)
\bigg[\overline\chi_L(P)\bar \sigma^z\bigg(\frac {2\pi^2}{|\vec p|}-\frac{\pi^2\rho^2}{z_+^2}{\cal P}_-U\bar z_+U^\dagger I_+^0\bigg)
\nonumber\\
&\times
({\rm cos}F(z, z_I)+i\,U^\dagger 2\sigma_{z\nu}\hat z_{I\nu}U\,{\rm sin}F(z, z_I)) S_0(z_-)\epsilon U)(U^\dagger\epsilon \chi_L(P)\bigg]
\end{align}
 $\Pi_{\pm}\rightarrow 1$ at large $z_I$, and the integrand is dominated by $S_0(z_I)$, which is seen to integrate to zero for $z=0$.
The apparent linear divergence vanishes. (\ref{ZMZ2}) is at most logarithmically divergent in $z_I$ which is cutoff by   the
mean  distance $L$ to the nearest neighbor,  hence of order $\kappa\,{\rm Log}(1/\kappa)$ and subleading in the diluteness expansion.

\begin{figure}[h!]
	\begin{center}
		\includegraphics[width=16cm]{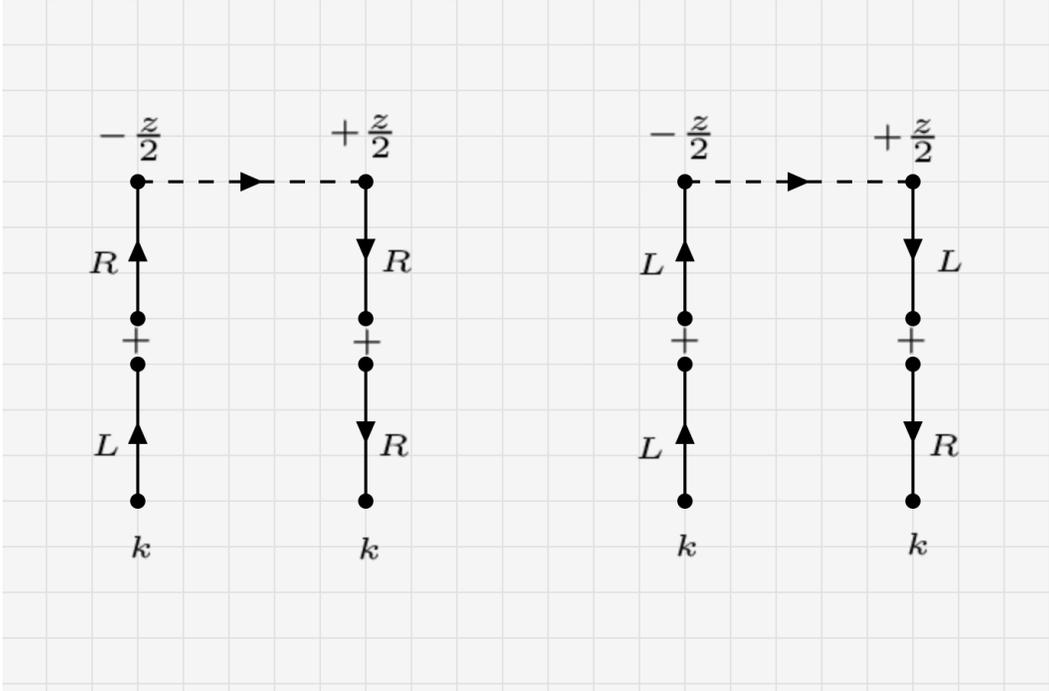}
		\caption{Non-vanishing small instanton contributions to twist-2 matching factor from mixing between non-zero-modes and zero-modes.
		}
		\label{fig-ZM}
	\end{center}
\end{figure}

\subsection{Comment on Current conservation}

Here we briefly comment on the conservation for the vector current $J^{\mu}=\bar\psi \gamma^{\mu}\psi$ and the normalization of the
PDF. At tree level, there is no matching effect, and the quasi-PDF normalizes to  $1$,

\begin{align}
\tilde f(z^2=zP^z=0)=1 \ .
\end{align}
However, the density expansion to leading order in $\sqrt{\kappa}$
 can differ from this canonical result. Additional charge renormalization is needed. Indeed, at $z=0$, the quasi-PDF reads

\begin{align}
\tilde f(0)=1+n_I\int d^4 z_I \tilde f^1(z_I)\ ,
\end{align}
with

\begin{align}
\tilde f^1(z_I)={\rm tr}{\cal P}_+ \sigma^z {\cal P}_+ {\rm tr}_c\left(\frac{1}{4}\bigg(\frac{1}{{\Pi_I}}-1\bigg)
-\frac{\rho^2}{8x_I^2}\frac{P^z}{N_c}({\cal P}_-\bar x_I I_+^1-x_I {\cal P}_+I_-^1)\frac{1}{\Pi_I}\right)+...
\end{align}
where the dots stand for the $z=0$ additional contributions listed in Appendix~\ref{ALL}.
Clearly, these integrals are non-zero, and a charge renormalization  $J^{\mu}\rightarrow Z J^{\mu}$ is needed, with
 $Z=1+\sqrt\kappa Z_1(\kappa)$. This will enforce  the condition $\tilde f(0)=1$. At the quasi-PDF level, this means $\tilde f\rightarrow \tilde f+\delta(1-x)\sqrt\kappa Z_1(\kappa)$, which will  normalize the quasi-PDF to 1.

\subsection{Small instanton contributions: polarized distributions}
\label{sec_polarized}


The instanton contributions to the polarized quark PDF are illustrated in Fig.~\ref{fig-ZM} with the result

\begin{align}\label{ZMX1}
&i\frac{n_{I+\bar I}}{2}\int d^4z_{I}\overline\chi_L(P)iP{\bar S}(P,\frac{z}{2}-z_I)  \sigma^z\bar \sigma^i[+\frac{z}{2},-\frac{z}{2}]_I{S}(-\frac{z}{2}-z_I,P)i\bar P\chi_R(P)\nonumber \\
-&i\frac{n_{I+\bar I}}{2}\int d^4z_{I}\overline\chi_R(P)i \bar P{S}(P,\frac{z}{2}-z_I)\bar \sigma^z\sigma^i[+\frac{z}{2},-\frac{z}{2}]_{I}{\bar S}(-\frac{z}{2}-z_I,P)i P\chi_L(P) \ ,
\end{align}
The anti-instanton contribution follows similarly and will be added at the end. We apply the same rules to unwound the matrix elements in (\ref{ZMX1}),
with the result ($z_\pm =\pm z/2-z_I$)

\begin{align}
\label{ZMX2}
&i\frac{n_{I+\bar I}}{2}\int d^4z_{I}\int dU \,e^{iP^zz}\bigg(\frac{4\pi^2\rho^3}{im\Pi_+^{\frac 32}}\bigg)\,\frac 1{\sqrt{\Pi_-}}\nonumber\\
&\times\bigg((\overline\chi_L(P)\epsilon U)(U^\dagger \epsilon \bar S_0(z_+))\sigma^z\bar\sigma^i
({\rm cos}F(z, z_I)+i\,U^\dagger 2\sigma_{z\nu}\hat z_{I\nu}U\,{\rm sin}F(z, z_I))\chi_R(P)\bigg)\nonumber\\
-&i\frac{n_{I+\bar I}}{2}\int d^4z_{I}\int dU \,e^{iP^zz}\bigg(\frac{4\pi^2\rho^3}{im\Pi_-^{\frac 32}}\bigg)\,\frac 1{\sqrt{\Pi_+}}\nonumber\\
&\times\bigg((\overline\chi_R(P)\bar \sigma^z\sigma^i
({\rm cos}F(z, z_I)+i\,U^\dagger 2\sigma_{z\nu}\hat z_{I\nu}U\,{\rm sin}F(z, z_I)) S_0(z_-)\epsilon U)(U^\dagger\epsilon \chi_L(P)\bigg)
\end{align}
with $S_0(z_-)=z_-/(2\pi^2z_-^4)$  and $\bar S_0(z_+)=z_+/(2\pi^2z_+^4)$. The reduction of the zero modes is given in Appendix~\ref{app_zero}.
The chirality flip is seen to follow from the mixing between the zero modes and the non-zero modes.

After carrying the averaging over the instanton moduli, only the  cos-term  contribution in (\ref{ZMX2})  survives.
The sin-term vanishes after the $z_I$-integration since $\sigma_{zz}=0$.
With this in mind and adding the contribution of the anti-instanton through $R\leftrightarrow L$ and conjugation
through the bar-assignment for $\sigma^\mu$, we obtain

\begin{align}
\label{ZMX3}
&e^{iP^zz}\frac {\kappa z}{\rho N_c}
\bigg({\cal J}_+(z)\,\overline\chi_L(P)\frac {\bar\sigma^z\sigma^z\bar\sigma^i}{2m}\chi_R(P)+{\cal J}_-(z)\,
\overline\chi_R(P)\frac{\bar\sigma^z\sigma^i\sigma^z}{2m}\chi_L(P)\bigg)\nonumber\\
+\,&e^{iP^zz}\frac {\kappa z}{\rho N_c}
\bigg({\cal J}_+(z)\,\overline\chi_R(P)\frac {\sigma^z\bar\sigma^z\sigma^i}{2m}\chi_L(P)+{\cal J}_-(z)\,
\overline\chi_L(P)\frac{\sigma^z\bar\sigma^i\bar\sigma^z}{2m}\chi_R(P)\bigg)
\end{align}
with the diluteness factor $\kappa=\pi^2\rho^4n_{I+\bar I}$ and

\bea
{\cal J}_\pm(z)=\frac 1{2\pi^2}\int \frac{d^4z_I}{z^4_I}\bigg(\frac 1{1+\frac{\rho^2}{z_I^2}}\bigg)^{\frac 32}
\bigg(\frac 1{1+\frac{\rho^2}{z\pm z_I^2}}\bigg)^{\frac 12}\,{\rm cos}F(z, z_I\pm z/2)\approx \frac 14{\rm Log}\bigg(\frac 1\kappa\bigg)+{\cal O}(z^2)\nonumber\\
\eea
The SIA contribution to the polarized quasi-PDF is of order $\kappa{\rm Log}(1/\kappa)$, but vanishes as  $z\rightarrow 0$ with no contribution to the polarized quark  PDF.

\section{Instanton contribution to  the quasi-PDF: gluon-exchange at one-loop level}~\label{sec_loop}
In the previous section, we have investigated the ``tree-level'' contribution to the quasi-PDF in the instanton background. The matching effect turns out to be trivial. In this section we investigate a sample gluon-exchange diagram in the instanton background that has non-negligible matching effect.
\subsection{Perturbative contribution in position space}

To streamline the short distance contribution to the one-gluon exchange in the instanton background, we first consider its perturbative analogue

\begin{align}\label{ONELOOP2}
&e^{-iP^zz}\int d^4x \,d^4y\,e^{-iP\cdot (x-y)}\,g_s^2C_F\,\Delta_0(x-y)\,\nonumber\\
&\times{\rm Tr}\bigg(\gamma_\mu S_0(x, \frac z2)\gamma^z[\frac{z}{2},-\frac{z}{2}]
S_0(-\frac z2, y)T^b\gamma_\mu \chi(P)\chi^\dagger(P)\bigg)\nonumber\\
\rightarrow &e^{-iP^zz}\int d^4x \,d^4y\,e^{-iP\cdot (x-y)}\,\frac{g_s^2C_F}{2(2\pi^2)^3}\,\frac 1{(x-y)^2}\,\frac 1{(x-\frac z2)^4}\,\frac 1{(y+\frac z2)^4}\nonumber\\
&\times{\rm Tr}\bigg(\gamma_\mu \gamma\cdot (x-\frac z2)\gamma^z[\frac{z}{2},-\frac{z}{2}]
\gamma\cdot (-\frac z2-y)\gamma_\mu\slashed P\bigg)
\end{align}
with $C_F=(N_c^2-1)/2$ and after setting the gauge link to 1 to probe the leading singularity.
The  integral in (\ref{ONELOOP2})  can be performed in position space by shifting  $x-z/2\rightarrow x+\frac{r}{2}$ and $y+z/2\rightarrow x-\frac{r}{2}$,
with the result

\begin{align}
\label{VAR1}
\frac{\alpha_sC_F}{4\pi^4}\int d^4x d^4r\frac{{\rm Tr}\gamma \cdot(x+\frac{r}{2})\gamma^z\gamma\cdot(x-\frac{r}{2})(\gamma^0-\gamma^z)}{(x+\frac{r}{2})^4(x-\frac{r}{2})^4(r+z)^2} e^{-iPz-iP\cdot r}
\end{align}
The integral in $x$ can be undone

\begin{align}
\label{VAR2}
\int d^4x \frac{{\rm Tr}\gamma \cdot(x+\frac{r}{2})\gamma^z\gamma\cdot(x-\frac{r}{2})(\gamma^0-\gamma^z)}{(x+\frac{r}{2})^4(x-\frac{r}{2})^4}=-8\pi^2\frac{r^0r^z-(r^0)^2+r_\perp^2}{r^4}
\end{align}
leading (\ref{VAR1}) in the form

\begin{align}
\label{VAR3}
-\frac{2\alpha_sC_Fe^{-iP^zz}}{\pi^2}\int d^4re^{-iP\cdot r}\frac{r^0r^z-(r^0)^2+r_\perp^2}{r^4(r+z)^2} \ .
\end{align}
In Euclidean space, one has $x^0=-ix^4$, thus one has to evaluate the integral

\begin{align}
\label{VAR4}
-\frac{2\alpha_sC_Fe^{-iP^zz}}{\pi^2}\int d^4re^{-iP\cdot r}\frac{-ir^4r^z+(r^4)^2+r_\perp^2}{r^4(r+z)^2} \ .
\end{align}
and then analytic continue to the value $P^2=0$. By using a Feynman parametrization, one has

\begin{align}
\label{VAR5}
-\frac{4\alpha_sC_Fe^{-iP^zz}}{\pi^2}\int_{0}^1 dt(1-t)e^{iP^ztz}\int d^4r e^{-iP\cdot r}\frac{-ir^4r^z+(r^4)^2+r_\perp^2}{(r^2+t(1-t)z^2)^3} \ .
\end{align}
The last integral can be evaluated

\begin{align}
\label{VAR6}
&\int d^4r e^{-iP\cdot r}\frac{-ir^4r^z+(r^4)^2+r_\perp^2}{(r^2+t(1-t)z^2)^3}\nonumber \\
&=-\bigg(i\frac{\partial}{\partial P^4}\frac{\partial}{\partial P^z}+\frac{\partial^2}{\partial (P^4)^2}+\frac{\partial^2}{\partial P_\perp^2}\bigg)\int d^4r e^{-iP\cdot r}\frac{1}{(r^2+t(1-t)z^2)^3} \ ,
\end{align}
which exhibits the desired singularity as $z^2\rightarrow 0$.

Now we can use the relation between the quasi PDF $\tilde f$ and the PDF $f$ in position space to extract the matching kernel $C$.

\begin{align}
\label{VAR7}
\tilde f(z^2,\lambda=zP^z)=\int_{0}^1 d\alpha C(z^2,\alpha)f(\alpha\lambda)
\end{align}
The iteration of (\ref{VAR7}) to one-loop gives

\begin{align}
\label{VAR8}
\tilde f^{(1)}(z^2,\lambda)-f^{(1)}(\lambda)=\int_{0}^1 d\alpha C^{(1)}(z^2,\alpha)e^{i\alpha \lambda}
\end{align}
Factorisation follows if (\ref{VAR8}) admits an IR safe solution kernel $C^{(1)}$. Note that the loop-expansion can be sought
either in perturbation theory, or semi-classically  as we already discussed at tree level. We now proceed to the semi-classical one-loop correction.

\subsection{Gluon exchange in the instanton background}

The gluon propagator in the instanton background is obtained using the decomposition $A_\mu=A_{I\mu}+a_\mu$ in the YM action
to quadratic order in the covariant background field gauge $D_\mu a_\mu=(\partial_\mu-iA_{I\mu}) a_\mu=0$,

\be
\frac 12 a_\mu\bigg(-D^2\delta_{\mu\nu}-2[D_\mu, D_\nu]+\bigg(\frac 1\alpha -1\bigg)D_\mu D_\nu\bigg)a_\nu
\ee
The instanton with collective variables $\xi_i=\lambda_i, \rho, Z_i$ is characterized by
$$[(N_c^2-1)-(N_c-2)^2]+1+4=4N_c$$
zero modes $\phi_{\mu}^i=\partial A_\mu(x, \xi_i)/\partial \xi_i$. As a result, the gauge-fixed gluon propagator follows from the inversion

\be
\label{ONELOOP0}
\bigg(-D^2\delta_{\mu\nu}-2[D_\mu, D_\nu]+\bigg(\frac 1\alpha -1\bigg)D_\mu D_\nu\bigg)\Delta_{\mu\nu}(x,y)=
\delta_{\mu\nu}\delta^4(x-y)-\sum_{i=1}^{4N_c}\phi_\mu^i(x)\phi_\nu^{i\dagger}(y)
\ee
The inversion can be obtained in closed form~\cite{Brown:1977eb}

\be
\label{ONELOOP00}
\Delta_{\mu\nu}(x,y)=g_{\mu\nu\rho\sigma}iD_\rho\Delta_2(x,y)iD_\sigma+(\alpha-1)iD_\mu\Delta_2(x,y)iD_\nu
\ee
in terms of the squared scalar propagator in the instanton background

\be
\Delta_2(x,y)=\int d^4z \Delta(x,z)\Delta(z,y)
\ee
with

\be
\label{ONELOOP000}
\Delta^{ab}(x,y)=\Delta_0(x-y)\bigg(1+\rho^2\frac {[Ux\bar y U^\dagger]}{x^2y^2}\bigg)^{ab}\frac 1{\sqrt{\Pi_x\Pi_y}}
\ee
and with the free scalar propagator $\Delta_0(x)=1/(2\pi x)^2$, and

\be
g_{\mu\nu\rho\sigma}=\bigg(\delta_{\mu\nu}\delta_{\rho\sigma}+\delta_{\mu\rho}\delta_{\nu\sigma}-\delta_{\mu\sigma}\delta_{\nu\rho}+\epsilon_{\mu\nu\rho\sigma}\bigg)
\ee

\subsection{Non-perturbative contribution in position space}


For the unpolarized PDF, the one-loop contribution in the instanton background
follows the tree level contributions in Fig.~\ref{fig-NZM} with an additional  gluon ring through the instanton,
that could attach to the pair of in-out quarks, the gauge-link or the quark-gauge link.  For an estimte, consider
the ring attached to the in-out quarks,

\begin{align}\label{ONELOOP1}
&\frac{n_{I+\bar I}}{2}\int d^4z_{I}\int d^4x \,d^4y\,\Delta^{ab}_{\mu\nu}(x,y)\,
\overline\chi_R(P)i\bar P{S}(P,z_I+x)g_sT^a\bar\sigma_\mu S(x, z_I+\frac z2)\nonumber\\
&\times \bar \sigma^z[\frac{z}{2},-\frac{z}{2}]_IS(z_I-\frac z2, y)g_sT^b\bar\sigma_\nu {S}(z_{I}+y,P)i\bar P\chi_R(P)\nonumber \\
+&\frac{n_{I+\bar I}}{2}\int d^4z_{I} \int d^4x \,d^4y\,\Delta^{ab}_{\mu\nu}(x,y)\,
\overline\chi_L(P)i P{\bar S}(P,z_{I}+x)g_sT^a\sigma_\mu \bar S(x, z_{ I}+\frac z2)\nonumber\\
&\times\sigma^z[\frac{z}{2},-\frac{z}{2}]_{ I}\bar S(z_{ I}-\frac z2, y)g_sT^b\sigma_\nu {S}(z_{ I}+y,P)iP\chi_L(P)
\end{align}
The anti-instanton contribution follows similarly and will be added at the end.

If we set the gauge link to 1,
the insertion of the gluon propagator (\ref{ONELOOP00}) simplifies, if we note that the
$(\alpha-1)$ gauge dependent insertion drops on mass-shell.  Also, the substitution

\begin{align}
\label{ONELOOPX2}
 \epsilon_{\mu\nu\rho\sigma}iD_\rho\Delta_2(x,y)iD_\sigma=&\frac 12\bigg(F_{\mu\nu}(x)\Delta_2(x,y)+\Delta_2(x,y)F_{\mu\nu}(y)\bigg)
\end{align}
using the self-duality of the the gauge field, vanishes after color tracing since $F_{\mu\nu}^{aa}(x)=0$.
Also, the contribution from the tensor part $(\delta_{\mu\rho}\delta_{\nu\sigma}-\delta_{\mu\sigma}\delta_{\nu\rho})$ vanishes for the
in-out quark on mass shell through an integration by parts.
With this in mind,  we now carry the remainder of the non-vanishing contributions using

\begin{align}
\label{ONELOOPX1}
iD_\rho\Delta_2(x,y)iD_\rho=\Delta(x,y)= \Delta_0(x-y)\bigg(\frac 1{\sqrt{\Pi_x\Pi_y}}+{\cal O}\bigg(\frac 1{N_c}\bigg)\bigg)
\end{align}
or more specifically

\be
\label{EST1}
\Delta^{ab}_{\mu\nu}(x,y)\rightarrow \delta^{ab}\delta_{\mu\nu}\,  \Delta_0(x-y)\bigg(\frac 1{\sqrt{\Pi_x\Pi_y}}+{\cal O}\bigg(\frac 1{N_c}\bigg)\bigg)
\ee
where the second contribution in (\ref{ONELOOP000}) is suppressed by $1/N_c$ after color averaging.
Similarly,

\begin{align}
\label{EST2}
&i\bar P S(P,z_I+x)\rightarrow {e^{-iP\cdot (z_I+x)}}\frac 1{\sqrt{\Pi_x}}+{\cal O}\bigg(\frac 1{N_c}\bigg)\nonumber\\
&S(P,z_I+y)i\bar P\rightarrow  {e^{+iP\cdot (z_I+y)}}\frac 1{\sqrt{\Pi_y}}+{\cal O}\bigg(\frac 1{N_c}\bigg)
\end{align}
for  the LSZ reduced quark propagator.
Throughout, we are using the short-hand notation

\be
\Pi_x=\frac 1{1+\frac{\rho^2}{(x-z_I)^2}}\qquad{\rm and}\qquad \Pi_y=\frac 1{1+\frac{\rho^2}{(y-z_I)^2}}
\ee
Inserting (\ref{EST1}-\ref{EST2}) into (\ref{ONELOOP1}), keeping only the
leading  fermion propagators in the loop in coordinate space to track the singular contribution, we obtain

\begin{align}\label{ONELOOP1X1}
&{n_{I+\bar I}}\,\frac{g_s^2C_F}{2(2\pi^2)^3}\,\int d^4z_{I} \,e^{-iP^zz}
\int d^4x \,d^4y\,e^{-iP\cdot (x-y)}\nonumber\\
&\times \frac 1{(x-y)^2}\,\frac 1{(x-\frac z2)^4}\,\frac 1{(y+\frac z2)^4}{\rm Tr}\bigg(\gamma_\mu \gamma\cdot (x-\frac z2)\gamma^z
\gamma\cdot (-\frac z2-y)\gamma_\mu\slashed P\bigg)\nonumber\\
&\times\bigg(\frac 1{\sqrt{\Pi_x}}\frac 1{\sqrt{\Pi_y}}\frac 1{\sqrt{\Pi_x\Pi_y}}-1\bigg)\bigg(\frac 1{\sqrt{\Pi_{x-z/2}}}\frac 1{\sqrt{\Pi_{y+z/2}}}\bigg)
\end{align}
The free contribution for $n_{I+\bar I}\rightarrow 1$ and $\rho\rightarrow 0$ is subtracted,
as it amounts to the one-loop  perturbative contribution evaluated earlier.
Using the same change of variables as in (\ref{VAR1}),  shifting
the $z_I$ integration variable,

\begin{align}
&x-z_I\rightarrow x+\frac r2 +\frac z2-z_I\rightarrow +\frac r2+\frac z2-z_I\equiv +(R_z-z_I)\nonumber\\
&y-z_I\rightarrow x-\frac r2 -\frac z2-z_I\rightarrow -\frac r2-\frac z2-z_I\equiv -(R_z+z_I)
\end{align}
with $R_z=(r+z)/2$, and

\begin{align}
&x-\frac z2-z_I\rightarrow x+\frac r2 -z_I\rightarrow +\frac r2-z_I\equiv +( R-z_I)\nonumber\\
&y+\frac z2-z_I\rightarrow x-\frac r2 -z_I\rightarrow -\frac r2-z_I\equiv -( R+z_I)
\end{align}
with $ R=r/2$, and carrying the x-integration as in (\ref{VAR1}-\ref{VAR4}) we obtain

\begin{align}
&-\frac{g_s^2C_Fn_{I+\bar I}\rho^4}{2\pi^3}e^{-iP^zz}
\int d^4re^{-iP\cdot r}\frac{-ir^4r^z+(r^4)^2+r_\perp^2}{r^4(r+z)^2} \,{\bf F}(R_z, R)
\end{align}
The  dimensionless instanton induced  form factor  in coordinate space is

\be
{\bf F}(R_z, R)=
\int \frac{d^4z_I}{\rho^4}
\bigg(\frac 1{\sqrt{\Pi_{+R_z}}}\frac 1{\sqrt{\Pi_{-R_z}}}\frac 1{\sqrt{\Pi_{+R_z}\Pi_{-R_z}}}-1\bigg)\frac 1{\sqrt{\Pi_{+ R}}}\frac 1{\sqrt{\Pi_{-R}}}
\ee
After a Feynman parametrization similar to (\ref{VAR5}), we finally get

\begin{align}
\label{XONELOOP1}
&-\frac{2g_s^2C_F\kappa}{\pi^5}
\int_{0}^1 dt(1-t)e^{iP^ztz}\int d^4r e^{-iP\cdot r}\frac{-ir^4r^z+(r^4)^2+r_\perp^2}{(r^2+t(1-t)z^2)^3}\,{\bf F}(R_z, R)
\end{align}
with the diluteness factor $\kappa=\pi^2\rho^4n_{I+\bar I}$.  For $r,z\rightarrow 0$, the integral in (\ref{XONELOOP1}) diverges
logarithmically as in the one-loop perturbative case (\ref{VAR6}) but with a non-perturbative pre-factor

\be
2\times\bigg(-\frac{2g_s^2C_F\kappa}{\pi^5}\bigg)\bigg(-\frac{2\pi^3}{\sqrt\kappa}\bigg)\,\int_{0}^1 dt(1-t)e^{iP^ztz}\int d^4r e^{-iP\cdot r}\frac{-ir^4r^z+(r^4)^2+r_\perp^2}{(r^2+t(1-t)z^2)^3}
\ee
with the extra factor of  2 accounting for the anti-instanton contribution, after using ($|z_I|\leq L$)

\begin{align}
{\bf F}(0,0)=&\int d^4z \bigg(\frac 1{({1+1/z^2})^2}-1\bigg)\frac 1{({1+1/z^2})}\approx-\frac{2\pi^3}{\sqrt\kappa}
\end{align}

\subsection{Matching kernel and trans-series}

To one-loop, the contribution of the instanton to the matching kernel amounts to a shift in the perturbative pre-factor by

\be
\label{TRANS}
-\frac{\alpha_s C_F}{\pi^2}\rightarrow -\frac{\alpha_s C_F}{\pi^2}\bigg(1-32\pi\sqrt\kappa\bigg)
\ee
The matching kernel coefficient develops into a trans-series~\cite{Dunne:2013ada} since at weak coupling $\sqrt\kappa\sim e^{-\pi/\alpha_s}$. As we noted earlier,
cooled lattice simulations of the QCD vacuum suggest $\kappa\approx 1/10$ (regime of chiral symmetry breaking with large size instantons \cite{Schafer:1996wv})
and $\kappa\approx 1$  (regime of short distance physics with small size instantons~\cite{Athenodorou:2018jwu}).  For both estimates, the small size
instanton contribution in  (\ref{TRANS})  is sizable in comparison to the perturbative one. It should be included when matching the quasi-PDF to the PDF in current lattice simulations.

\section{Conclusions}
\label{sec_conclusions}

Cooled lattice gauge configurations display strongly inhomogeneous instanton and anti-instanton configurations. The dilute QCD instanton vacuum
capture the essentials of these configurations. The small size instantons and anti-instantons involve strong gauge fields that contribute non-perturbatively
to short distance physics without modifying the essentials of the UV behavior. Using the SIA, we have
shown how the instantons contribute sizably to the quark PDF in leading order. More importantly, although the matching effects is trivial at tree-level, we have found that they are part of a trans-series expansion of the matching kernel, that is dominant at one-loop. They should be considered in the current lattice extraction of the PDF$^\prime$s from their quasi-PDF counterparts.

We should emphasise that only the one-gluon contribution diagram has been considered, which is sufficient for our purpose of estimating the size of the effect. More detailed calculation with all diagrams included shall be provided somewhere else together with a calculation of the non-perturbative renormalization factors.

\vskip 1cm
{\bf Acknowledgements}
This work is supported by the Office of Science, U.S. Department of Energy under Contract No. DE-FG-88ER40388.

\appendix

\section{Conventions}\label{app_defs}

Following~\cite{Balitsky:1993ki,Moch:1996bs,Vandoren:2008xg},
we use the short hand matrix-valued notation
 $x\equiv \sigma_\mu x^\mu$ and $\overline{x}\equiv \overline\sigma_\mu x^\nu$, with
the covariantized Pauli matrices in Euclidean and Minkowski space defined as

\begin{eqnarray} \label{eqn_4dsigma}
{\rm Euclidean}:&&\qquad \sigma_\mu=(1,-i\vec\sigma)\qquad \overline\sigma_\mu=(1, +i\vec\sigma)\qquad
\sigma_\mu\overline\sigma_\nu+\sigma_\nu\overline\sigma_\mu=2\eta_{\mu\nu}\nonumber\\
{\rm Minkowski}:&&\qquad \sigma_\mu=(1,-\vec\sigma)\qquad \overline\sigma_\mu=(1, +\vec\sigma)\qquad
\sigma_\mu\overline\sigma_\nu+\sigma_\nu\overline\sigma_\mu=2g_{\mu\nu}\nonumber
\end{eqnarray}
with metric $g^{\mu\nu}=(+,-,-,-)$, $\eta^{\mu\nu}=\delta^{\mu\nu}$, and satisfying the identities

\begin{equation}
\sigma^\mu\overline{\sigma}^\nu-\sigma^\nu\overline{\sigma}^\mu=2i\overline{\eta}^{a\mu\nu}\tau^a\qquad\qquad
\overline\sigma^\mu{\sigma}^\nu-\overline\sigma^\nu{\sigma}^\mu=2i{\eta}^{a\mu\nu}\tau^a
\end{equation}
with the $\eta$-tHooft symbol.
The spinor indices are $\alpha,\beta=1,2$, and the color indices are  $i,j=1,2, ... N_c$.






Our  conventions for the $\gamma^5$ matrix in Euclidean and Minkowski space are respectively

\be
\gamma_E^5=
\begin{pmatrix}
-1& 0\\
0 &1
\end{pmatrix}\qquad\qquad
\gamma_M^5=
\begin{pmatrix}
1& 0\\
0 &-1
\end{pmatrix}
\ee
In Weyl notations, the Euclidean Dirac spinor reads

\begin{eqnarray}
\label{NOTE1}
\Psi(x)&=&
\begin{pmatrix}
K^i_\alpha(x)\\
\phi^\alpha_i(x) \\
\end{pmatrix}
\qquad
\Psi^\dagger(x)=({K}_i^{\dagger\alpha}(x),\phi_\alpha^{\dagger i}(x))\equiv   (\overline{\phi}^\alpha_i(x), \overline{K}^i_\alpha(x))
\end{eqnarray}
The Euclidean fermionic action splits into left $K$ and right $\phi$ copies

\be
\overline K\sigma\cdot (\partial-igA)K+\overline \phi\,\overline\sigma\cdot(\partial -igA)\phi
\ee
with  $\overline K=\phi^\dagger$ and  $\overline\phi=K^\dagger$ using

\be
\gamma_E^\mu=
\begin{pmatrix}
0& \bar\sigma_E^\mu\\
\sigma_E^\mu &0
\end{pmatrix}\qquad\qquad
\gamma_M^\mu=
\begin{pmatrix}
0& \bar\sigma_M^\mu\\
\sigma_M^\mu &0
\end{pmatrix}
\ee

\section{Fermionic zero modes }\label{app_zero}

The instanton admits a left-handed zero mode $K^i_\alpha(x)$  satisfying $\sigma\cdot D\,K=0$,
and the anti-instanton a right-handed zero mode $\phi^\alpha_i(x)$ satisfying $\overline\sigma\cdot D\,\phi=0$,
which are eigenstates of $(1\pm\gamma_E^5)/2$, and conjugate of each other.
In terms of the Euclidean Weyl spinors,  the instanton zero mode  and its
conjugate are


\begin{eqnarray}
\label{LR}
K^i_\alpha(x)&=&\frac{\rho^{\frac 32}}{\pi x^4}\frac{(\overline{x}\epsilon U)^i_\alpha}{\Pi_x^{\frac 32}}
=\frac{2\pi\rho^{\frac 32}}{\Pi_x^{\frac 32}}\,(\overline{S}_0(x)\epsilon U)^i_\alpha
\nonumber\\
K^{\dagger\alpha}_i(x)&=&\frac{\rho^{\frac 32}}{\pi x^4}\frac{(U^\dagger\epsilon x)^\alpha_i}{x^4\Pi_x^{\frac 32}}
=\frac{2\pi\rho^{\frac 32}}{\Pi_x^{\frac 32}}\,(U^\dagger\epsilon {S}_0(x))_i^\alpha\equiv {\overline\phi}^\alpha_i(x)
\end{eqnarray}
Here $\epsilon$ is the antisymmetric spin 2-tensor with the normalization
$\epsilon_{\alpha\sigma}\epsilon^{\sigma\beta}=\delta_\alpha^\beta$, and

\begin{equation}
\Pi_x=1+\frac{\rho^2}{x^2}\qquad S_0(x)=\frac{x}{2\pi^2x^4}\qquad  \overline{S}_0(x)=\frac{\overline x}{2\pi^2x^4}
\end{equation}
with $S_0(x)$ the free massless quark propagator.
The zero modes  are normalized to $\rho$,

\bea
\int d^4x\, K^\dagger(x) K(x)=\rho\qquad \qquad \int d^4x \,\phi^\dagger(x)\phi(x)=\rho
\eea

For the free Dirac spinors we will use the notation $\chi(k)=\chi_R(k)+\chi_L(k)$  (with Minkowski labeling) as  the sum of  free Weyl spinors,  that  satisfy

\begin{eqnarray}
\label{FREEX}
\slashed k\chi(k) =\slashed  k
\begin{pmatrix}
\chi_R(k) \\
\chi_L(k) \\
\end{pmatrix}
=
\begin{pmatrix}
0&\overline{k} \\
k& 0\\
\end{pmatrix}
\begin{pmatrix}
\chi_R(k)\\
\chi_L(k) \\
\end{pmatrix}
=
\begin{pmatrix}
\overline{k}\chi_L(k) \\
k\chi_R(k)\\
\end{pmatrix}
=0
\end{eqnarray}
with the free-wave ortho-normalizations

\begin{eqnarray}
\chi_{L,R}(k)\chi_{L,R}^\dagger(k)=k,\overline k\qquad \qquad \chi^\dagger_{L,R}(k)\chi_{R,L}(k)=0
\end{eqnarray}

\section{Details of the large time limit}~\label{app_large}

In this Appendix we show how the large time limit in (\ref{T0}) can be taken. One can see that (\ref{T0}) consists of several contributions as in (\ref{BROWNCHIRAL}), with the last and more involved contribution

\begin{align}
\label{T1}
\int d^3\vec{y} e^{-i\vec{p}\cdot \vec{y}}\frac{(x\bar y-y^2)}{(x-y)^2y^2}=\int d^3\vec{y} e^{-i\vec{p}\cdot \vec{y}}\frac{(x\bar y)}{(x-y)^2y^2}-\int d^3\vec{y} e^{-i\vec{p}\cdot \vec{y}}\frac{1}{(x-y)^2} \ ,
\end{align}
with $y=(-T,\vec{y})$ and  $x=(T_1,\vec{x})$.  Note that  $1/\Pi_y^{\frac 12}$ asymptotes 1 at large $T$.
The second contribution in (\ref{T1}) is the  free scalar  propagator and is readily integrated as detailed in the siummary integrals  below.
The first contribution can be integrated through a Feynman parametrization followed by a change of variable
$\vec{y}=t\vec{x}+\vec{z}$,  to give

\begin{align}
\label{T2}
&x\int_{0}^1 dt\int d^3\vec{z} e^{-i\vec{p}\cdot \vec{z}-it\vec{p}\cdot \vec{x}}
\frac{-T+it\vec{\sigma}\cdot \vec{x}+i\vec{\sigma}\cdot \vec{z} }{(\vec{z}^2+(T+tT_1)^2+(t-t^2)\vec{x}^2)^2} \nonumber\\
&=\pi^2x\int_{0}^1 dt(-T+it\vec{\sigma}\cdot\vec{x}-\sigma\cdot \vec{\nabla}_{\vec{p}})\frac{e^{-|\vec{p}|\sqrt{(T+tT_1)^2+(t-t^2)\vec{x}^2}}}{\sqrt{(T+tT_1)^2+(t-t^2)\vec{x}^2}} \nonumber\\
&\rightarrow
e^{-|T||\vec{p}|}\bigg(\frac{-|\vec{p}|+\vec{\sigma}\cdot \vec{p}}{|\vec{p}|}\pi^2 x\int_{0}^{1} dt e^{-t|\vec{p}|T_1-it\vec{p}\cdot\vec{x}}+{\cal O}\left( \frac{1}{|T|}\right)\bigg) \ .
\end{align}
Similarly,  consider the left reduction of $\bar S$

\be
\lim_{T\rightarrow \infty}\int d^3\vec{x}\bar S(T,\vec{x};y)e^{i\vec{p}\cdot \vec{x}}
\ee
which consists also of several contributions in (\ref{BROWNCHIRAL}). The last and involved contribution

\begin{align}
\int d^3 \vec{y}e^{i\vec{p}\cdot \vec{y}}\frac{y\bar \sigma^{\mu}(y-x)}{(y^2)^2(y-x)^2} \ ,
\end{align}
can be unwound and evaluated through a Feynman parametrization

\begin{align}
\label{T3}
&2\int (1-t)dt \int d^3 \vec{z}e^{i\vec{p}\cdot \vec{z}+it\vec{p}\cdot \vec{x}}\frac{(T-i\vec{\sigma} \cdot \vec{z}-it\vec{\sigma} \cdot \vec{x})\bar \sigma^{\mu}(T-T_1-i\vec{\sigma} \cdot \vec{z}+i(1-t)\vec{\sigma} \cdot \vec{x})}{(\vec{z}^2+(T-tT_1)^2+(t-t^2)\vec{x}^2)^3}\nonumber\\
&\rightarrow e^{-|T||\vec{p}|}\frac{\pi^2 |\vec{p}|}{2}(1+\frac{\vec{\sigma}\cdot \vec{p}}{|\vec{p}|})\bar \sigma^{\mu}(1+\frac{\vec{\sigma}\cdot \vec{p}}{|\vec{p}|})\int_{0}^{1}dt(1-t)e^{T_1t|\vec{p}|+it\vec{p}\cdot\vec{x}}
\end{align}
at large $T$.  Finally, we consider last

\begin{align}
\int d^3 \vec{y}e^{i\vec{p}\cdot \vec{y}}\frac{(\bar y-\bar x)Uy\bar xU^{\dagger}}{(y-x)^4y^2} \ .
\end{align}
A rerun of the preceding arguments give

\begin{align}
\label{T4}
e^{-|T||\vec{p}|}\frac{\pi^2 |\vec{p}|}{2}(1-\frac{\vec{\sigma}\cdot \vec{p}}{|\vec{p}|})U(1+\frac{\vec{\sigma}\cdot \vec{p}}{|\vec{p}|})\bar xU^{\dagger}\int_{0}^{1}dtte^{T_1t|\vec{p}|+it\vec{p}\cdot\vec{x}}
\end{align}
at large $T$.

We now note that the free propagators  at large $T$ read

\begin{align}
\label{T5}
\int d^3\vec{y}e^{i\vec{p}\cdot\vec{y}}\frac{\bar y -\bar x}{2\pi^2(y-x)^4}\rightarrow&  \frac{e^{-|T||\vec{p}|}}{2}(1-\frac{\vec{\sigma}\cdot \vec{p}}{|\vec{p}|})e^{T_1|\vec{p}|+i\vec{p}\cdot\vec{x}}\\
\int d^3\vec{y}e^{-i\vec{p}\cdot\vec{y}}\frac{\bar x -\bar y}{2\pi^2(y-x)^4}\rightarrow&  \frac{e^{-|T||\vec{p}|}}{2} (1-\frac{\vec{\sigma}\cdot \vec{p}}{|\vec{p}|})e^{-T_1|\vec{p}|-i\vec{p}\cdot\vec{x}}\\
\int d^3\vec{y}e^{-i\vec{p}\cdot\vec{y}}\frac{1}{(x-y)^2}\rightarrow& \frac{2\pi^2}{|\vec{p}|}e^{-T_1|\vec{p}|-i\vec{p}\cdot\vec{x}}
\end{align}

qpdf1 unpolarized
\section{All terms for the  quasi-PDF}~\label{ALL}

Here, we list all the contributions to the quasi-PDF discussed in  section~\ref{unpolarized} for completeness.
All the contributions to the quasi-PDF at $z=0$ are

\begin{align}
&A={\rm tr}{\cal P}_+ \sigma^z {\cal P}_+ {\rm tr}_c\left(\frac{1}{4}(\frac{1}{\sqrt{\Pi_I}}-1)^2-\frac{\rho^2}{8x_I^2N_c}P^z({\cal P}_-\bar x_I I_+^1-x_I {\cal P}_+I_-^1)\frac{1}{\Pi_I}\right)\\
&B=-{\rm tr}{\cal P}_+\sigma^z\bar \sigma_{\mu}{\rm tr}_c x_I\bar \sigma_{\mu}x_I{\cal P}_+ I_- \frac{\rho^2}{8N_cx_I^4 \Pi_I^2}\\
&C={\rm tr}{\cal P}_+\sigma^z\bar \sigma_{\mu}{\rm tr}_c x_I\bar \sigma_{\mu}\frac{\rho^2}{4N_cx_I^4p^z\Pi_I^2} \\
&D={\rm tr}{\cal P}_+\sigma^z\bar \sigma_{\mu}{\rm tr}_c {\cal P}_-\bar x_I x_I\sigma^{\mu}\frac{\rho^4}{8N_cx_I^6 \Pi_I^2} I_+^1\\
&E=-{\rm tr}\bar \sigma^{\mu}\sigma^z {\cal P}_+{\rm tr}_c {\cal P}_-\sigma_{\mu}{\cal P}_-\bar x_I p^z\frac{\rho^2}{8N_cx_I^2\Pi_x}I_+ \\
&F=-{\rm tr}\bar \sigma^{\mu}\sigma^z\bar \sigma_{\mu'}{\rm tr}_c {\cal P}_-\sigma_{\mu}{\cal P}_-\bar x_I x_I \sigma^{\mu'}\frac{\rho^4}{16N_cx_I^6\Pi_I^2}I_+ \ .
\end{align}
Similarly, the full five contributions to the quasi-PDF at $z\neq 0$ are

\begin{align}
&B=-{\rm tr}{\cal P}_+\sigma^z\bar \sigma_{\mu}{\rm tr}_c x^+_I\bar \sigma_{\mu}x^-_I{\cal P}_+ I_- \frac{\rho^2}{8N_c(x^-_I)^4 \sqrt{\Pi_I^+}(\Pi_I^-)^{\frac{3}{2}}}\\
&C={\rm tr}{\cal P}_+\sigma^z\bar \sigma_{\mu}{\rm tr}_c x^-_I\bar \sigma_{\mu}\frac{\rho^2}{4N_c(x^-_I)^4 \sqrt{\Pi_I^+}(\Pi_I^-)^{\frac{3}{2}}} \\
&D={\rm tr}{\cal P}_+\sigma^z\bar \sigma_{\mu}{\rm tr}_c {\cal P}_-\bar x^+_I x^-_I\sigma^{\mu}\frac{\rho^4}{8N_c(x^+_I)^2(x^-_I)^4\sqrt{\Pi_I^+}(\Pi_I^-)^{\frac{3}{2}}} I_+^1\\
&E=-{\rm tr}\bar \sigma^{\mu}\sigma^z {\cal P}_+{\rm tr}_c {\cal P}_-\sigma_{\mu}{\cal P}_-\bar x^+_I p^z\frac{\rho^2}{8N_c(x^+_I)^2\sqrt{\Pi_I^+\Pi_I^-}}I_+ \\
&F=-{\rm tr}\bar \sigma^{\mu}\sigma^z\bar \sigma_{\mu'}{\rm tr}_c {\cal P}_-\sigma_{\mu}{\cal P}_-\bar x^+_I x^-_I \sigma^{\mu'}\frac{\rho^4}{16N_c(x^+_I)^2(x^-_I)^4\sqrt{\Pi_I^+}(\Pi_I^-)^{\frac{3}{2}}}I_+ \ .
\end{align}
In section~\ref{unpolarized} and
for the purpose of estimating the speed of convergence of the quasi-PDF to the PDF in the large $P^z$ limit,
we have only considered the contribution corresponding to the terms in $A$.

\bibliographystyle{apsrev4-1}
\bibliography{ILAMET1230}

\begin{thebibliography}{26}%
\makeatletter
\providecommand \@ifxundefined [1]{%
 \@ifx{#1\undefined}
}%
\providecommand \@ifnum [1]{%
 \ifnum #1\expandafter \@firstoftwo
 \else \expandafter \@secondoftwo
 \fi
}%
\providecommand \@ifx [1]{%
 \ifx #1\expandafter \@firstoftwo
 \else \expandafter \@secondoftwo
 \fi
}%
\providecommand \natexlab [1]{#1}%
\providecommand \enquote  [1]{``#1''}%
\providecommand \bibnamefont  [1]{#1}%
\providecommand \bibfnamefont [1]{#1}%
\providecommand \citenamefont [1]{#1}%
\providecommand \href@noop [0]{\@secondoftwo}%
\providecommand \href [0]{\begingroup \@sanitize@url \@href}%
\providecommand \@href[1]{\@@startlink{#1}\@@href}%
\providecommand \@@href[1]{\endgroup#1\@@endlink}%
\providecommand \@sanitize@url [0]{\catcode `\\12\catcode `\$12\catcode
  `\&12\catcode `\#12\catcode `\^12\catcode `\_12\catcode `\%12\relax}%
\providecommand \@@startlink[1]{}%
\providecommand \@@endlink[0]{}%
\providecommand \url  [0]{\begingroup\@sanitize@url \@url }%
\providecommand \@url [1]{\endgroup\@href {#1}{\urlprefix }}%
\providecommand \urlprefix  [0]{URL }%
\providecommand \Eprint [0]{\href }%
\providecommand \doibase [0]{http://dx.doi.org/}%
\providecommand \selectlanguage [0]{\@gobble}%
\providecommand \bibinfo  [0]{\@secondoftwo}%
\providecommand \bibfield  [0]{\@secondoftwo}%
\providecommand \translation [1]{[#1]}%
\providecommand \BibitemOpen [0]{}%
\providecommand \bibitemStop [0]{}%
\providecommand \bibitemNoStop [0]{.\EOS\space}%
\providecommand \EOS [0]{\spacefactor3000\relax}%
\providecommand \BibitemShut  [1]{\csname bibitem#1\endcsname}%
\let\auto@bib@innerbib\@empty
\bibitem [{\citenamefont {Ji}(2013)}]{Ji:2013dva}%
  \BibitemOpen
  \bibfield  {author} {\bibinfo {author} {\bibfnamefont {X.}~\bibnamefont
  {Ji}},\ }\href {\doibase 10.1103/PhysRevLett.110.262002} {\bibfield
  {journal} {\bibinfo  {journal} {Phys. Rev. Lett.}\ }\textbf {\bibinfo
  {volume} {110}},\ \bibinfo {pages} {262002} (\bibinfo {year} {2013})},\
  \Eprint {http://arxiv.org/abs/1305.1539} {arXiv:1305.1539 [hep-ph]}
  \BibitemShut {NoStop}%
\bibitem [{\citenamefont {Zhang}\ \emph {et~al.}(2017)\citenamefont {Zhang},
  \citenamefont {Chen}, \citenamefont {Ji}, \citenamefont {Jin},\ and\
  \citenamefont {Lin}}]{Zhang:2017bzy}%
  \BibitemOpen
  \bibfield  {author} {\bibinfo {author} {\bibfnamefont {J.-H.}\ \bibnamefont
  {Zhang}}, \bibinfo {author} {\bibfnamefont {J.-W.}\ \bibnamefont {Chen}},
  \bibinfo {author} {\bibfnamefont {X.}~\bibnamefont {Ji}}, \bibinfo {author}
  {\bibfnamefont {L.}~\bibnamefont {Jin}}, \ and\ \bibinfo {author}
  {\bibfnamefont {H.-W.}\ \bibnamefont {Lin}},\ }\href {\doibase
  10.1103/PhysRevD.95.094514} {\bibfield  {journal} {\bibinfo  {journal} {Phys.
  Rev. D}\ }\textbf {\bibinfo {volume} {95}},\ \bibinfo {pages} {094514}
  (\bibinfo {year} {2017})},\ \Eprint {http://arxiv.org/abs/1702.00008}
  {arXiv:1702.00008 [hep-lat]} \BibitemShut {NoStop}%
\bibitem [{\citenamefont {Ji}\ \emph {et~al.}(2015)\citenamefont {Ji},
  \citenamefont {Sch\"afer}, \citenamefont {Xiong},\ and\ \citenamefont
  {Zhang}}]{Ji:2015qla}%
  \BibitemOpen
  \bibfield  {author} {\bibinfo {author} {\bibfnamefont {X.}~\bibnamefont
  {Ji}}, \bibinfo {author} {\bibfnamefont {A.}~\bibnamefont {Sch\"afer}},
  \bibinfo {author} {\bibfnamefont {X.}~\bibnamefont {Xiong}}, \ and\ \bibinfo
  {author} {\bibfnamefont {J.-H.}\ \bibnamefont {Zhang}},\ }\href {\doibase
  10.1103/PhysRevD.92.014039} {\bibfield  {journal} {\bibinfo  {journal} {Phys.
  Rev. D}\ }\textbf {\bibinfo {volume} {92}},\ \bibinfo {pages} {014039}
  (\bibinfo {year} {2015})},\ \Eprint {http://arxiv.org/abs/1506.00248}
  {arXiv:1506.00248 [hep-ph]} \BibitemShut {NoStop}%
\bibitem [{\citenamefont {Bali}\ \emph {et~al.}(2018)\citenamefont {Bali},
  \citenamefont {Braun}, \citenamefont {Gl\"a\ss{}le}, \citenamefont
  {G\"ockeler}, \citenamefont {Gruber}, \citenamefont {Hutzler}, \citenamefont
  {Korcyl}, \citenamefont {Sch\"afer}, \citenamefont {Wein},\ and\
  \citenamefont {Zhang}}]{Bali:2018spj}%
  \BibitemOpen
  \bibfield  {author} {\bibinfo {author} {\bibfnamefont {G.~S.}\ \bibnamefont
  {Bali}}, \bibinfo {author} {\bibfnamefont {V.~M.}\ \bibnamefont {Braun}},
  \bibinfo {author} {\bibfnamefont {B.}~\bibnamefont {Gl\"a\ss{}le}}, \bibinfo
  {author} {\bibfnamefont {M.}~\bibnamefont {G\"ockeler}}, \bibinfo {author}
  {\bibfnamefont {M.}~\bibnamefont {Gruber}}, \bibinfo {author} {\bibfnamefont
  {F.}~\bibnamefont {Hutzler}}, \bibinfo {author} {\bibfnamefont
  {P.}~\bibnamefont {Korcyl}}, \bibinfo {author} {\bibfnamefont
  {A.}~\bibnamefont {Sch\"afer}}, \bibinfo {author} {\bibfnamefont
  {P.}~\bibnamefont {Wein}}, \ and\ \bibinfo {author} {\bibfnamefont {J.-H.}\
  \bibnamefont {Zhang}},\ }\href {\doibase 10.1103/PhysRevD.98.094507}
  {\bibfield  {journal} {\bibinfo  {journal} {Phys. Rev. D}\ }\textbf {\bibinfo
  {volume} {98}},\ \bibinfo {pages} {094507} (\bibinfo {year} {2018})},\
  \Eprint {http://arxiv.org/abs/1807.06671} {arXiv:1807.06671 [hep-lat]}
  \BibitemShut {NoStop}%
\bibitem [{\citenamefont {Alexandrou}\ \emph {et~al.}(2018)\citenamefont
  {Alexandrou}, \citenamefont {Cichy}, \citenamefont {Constantinou},
  \citenamefont {Jansen}, \citenamefont {Scapellato},\ and\ \citenamefont
  {Steffens}}]{Alexandrou:2018eet}%
  \BibitemOpen
  \bibfield  {author} {\bibinfo {author} {\bibfnamefont {C.}~\bibnamefont
  {Alexandrou}}, \bibinfo {author} {\bibfnamefont {K.}~\bibnamefont {Cichy}},
  \bibinfo {author} {\bibfnamefont {M.}~\bibnamefont {Constantinou}}, \bibinfo
  {author} {\bibfnamefont {K.}~\bibnamefont {Jansen}}, \bibinfo {author}
  {\bibfnamefont {A.}~\bibnamefont {Scapellato}}, \ and\ \bibinfo {author}
  {\bibfnamefont {F.}~\bibnamefont {Steffens}},\ }\href {\doibase
  10.1103/PhysRevD.98.091503} {\bibfield  {journal} {\bibinfo  {journal} {Phys.
  Rev. D}\ }\textbf {\bibinfo {volume} {98}},\ \bibinfo {pages} {091503}
  (\bibinfo {year} {2018})},\ \Eprint {http://arxiv.org/abs/1807.00232}
  {arXiv:1807.00232 [hep-lat]} \BibitemShut {NoStop}%
\bibitem [{\citenamefont {Izubuchi}\ \emph {et~al.}(2019)\citenamefont
  {Izubuchi}, \citenamefont {Jin}, \citenamefont {Kallidonis}, \citenamefont
  {Karthik}, \citenamefont {Mukherjee}, \citenamefont {Petreczky},
  \citenamefont {Shugert},\ and\ \citenamefont {Syritsyn}}]{Izubuchi:2019lyk}%
  \BibitemOpen
  \bibfield  {author} {\bibinfo {author} {\bibfnamefont {T.}~\bibnamefont
  {Izubuchi}}, \bibinfo {author} {\bibfnamefont {L.}~\bibnamefont {Jin}},
  \bibinfo {author} {\bibfnamefont {C.}~\bibnamefont {Kallidonis}}, \bibinfo
  {author} {\bibfnamefont {N.}~\bibnamefont {Karthik}}, \bibinfo {author}
  {\bibfnamefont {S.}~\bibnamefont {Mukherjee}}, \bibinfo {author}
  {\bibfnamefont {P.}~\bibnamefont {Petreczky}}, \bibinfo {author}
  {\bibfnamefont {C.}~\bibnamefont {Shugert}}, \ and\ \bibinfo {author}
  {\bibfnamefont {S.}~\bibnamefont {Syritsyn}},\ }\href {\doibase
  10.1103/PhysRevD.100.034516} {\bibfield  {journal} {\bibinfo  {journal}
  {Phys. Rev. D}\ }\textbf {\bibinfo {volume} {100}},\ \bibinfo {pages}
  {034516} (\bibinfo {year} {2019})},\ \Eprint
  {http://arxiv.org/abs/1905.06349} {arXiv:1905.06349 [hep-lat]} \BibitemShut
  {NoStop}%
\bibitem [{\citenamefont {Izubuchi}\ \emph {et~al.}(2018)\citenamefont
  {Izubuchi}, \citenamefont {Ji}, \citenamefont {Jin}, \citenamefont
  {Stewart},\ and\ \citenamefont {Zhao}}]{Izubuchi:2018srq}%
  \BibitemOpen
  \bibfield  {author} {\bibinfo {author} {\bibfnamefont {T.}~\bibnamefont
  {Izubuchi}}, \bibinfo {author} {\bibfnamefont {X.}~\bibnamefont {Ji}},
  \bibinfo {author} {\bibfnamefont {L.}~\bibnamefont {Jin}}, \bibinfo {author}
  {\bibfnamefont {I.~W.}\ \bibnamefont {Stewart}}, \ and\ \bibinfo {author}
  {\bibfnamefont {Y.}~\bibnamefont {Zhao}},\ }\href {\doibase
  10.1103/PhysRevD.98.056004} {\bibfield  {journal} {\bibinfo  {journal} {Phys.
  Rev. D}\ }\textbf {\bibinfo {volume} {98}},\ \bibinfo {pages} {056004}
  (\bibinfo {year} {2018})},\ \Eprint {http://arxiv.org/abs/1801.03917}
  {arXiv:1801.03917 [hep-ph]} \BibitemShut {NoStop}%
\bibitem [{\citenamefont {Ji}\ \emph {et~al.}(2019)\citenamefont {Ji},
  \citenamefont {Liu},\ and\ \citenamefont {Zahed}}]{Ji:2018waw}%
  \BibitemOpen
  \bibfield  {author} {\bibinfo {author} {\bibfnamefont {X.}~\bibnamefont
  {Ji}}, \bibinfo {author} {\bibfnamefont {Y.}~\bibnamefont {Liu}}, \ and\
  \bibinfo {author} {\bibfnamefont {I.}~\bibnamefont {Zahed}},\ }\href
  {\doibase 10.1103/PhysRevD.99.054008} {\bibfield  {journal} {\bibinfo
  {journal} {Phys. Rev. D}\ }\textbf {\bibinfo {volume} {99}},\ \bibinfo
  {pages} {054008} (\bibinfo {year} {2019})},\ \Eprint
  {http://arxiv.org/abs/1807.07528} {arXiv:1807.07528 [hep-ph]} \BibitemShut
  {NoStop}%
\bibitem [{\citenamefont {Radyushkin}(2017)}]{Radyushkin:2017gjd}%
  \BibitemOpen
  \bibfield  {author} {\bibinfo {author} {\bibfnamefont {A.~V.}\ \bibnamefont
  {Radyushkin}},\ }\href {\doibase 10.1103/PhysRevD.95.056020} {\bibfield
  {journal} {\bibinfo  {journal} {Phys. Rev. D}\ }\textbf {\bibinfo {volume}
  {95}},\ \bibinfo {pages} {056020} (\bibinfo {year} {2017})},\ \Eprint
  {http://arxiv.org/abs/1701.02688} {arXiv:1701.02688 [hep-ph]} \BibitemShut
  {NoStop}%
\bibitem [{\citenamefont {Ma}\ and\ \citenamefont {Qiu}(2018)}]{Ma:2014jla}%
  \BibitemOpen
  \bibfield  {author} {\bibinfo {author} {\bibfnamefont {Y.-Q.}\ \bibnamefont
  {Ma}}\ and\ \bibinfo {author} {\bibfnamefont {J.-W.}\ \bibnamefont {Qiu}},\
  }\href {\doibase 10.1103/PhysRevD.98.074021} {\bibfield  {journal} {\bibinfo
  {journal} {Phys. Rev. D}\ }\textbf {\bibinfo {volume} {98}},\ \bibinfo
  {pages} {074021} (\bibinfo {year} {2018})},\ \Eprint
  {http://arxiv.org/abs/1404.6860} {arXiv:1404.6860 [hep-ph]} \BibitemShut
  {NoStop}%
\bibitem [{\citenamefont {Chu}\ \emph {et~al.}(1994)\citenamefont {Chu},
  \citenamefont {Grandy}, \citenamefont {Huang},\ and\ \citenamefont
  {Negele}}]{Chu:1994vi}%
  \BibitemOpen
  \bibfield  {author} {\bibinfo {author} {\bibfnamefont {M.~C.}\ \bibnamefont
  {Chu}}, \bibinfo {author} {\bibfnamefont {J.~M.}\ \bibnamefont {Grandy}},
  \bibinfo {author} {\bibfnamefont {S.}~\bibnamefont {Huang}}, \ and\ \bibinfo
  {author} {\bibfnamefont {J.~W.}\ \bibnamefont {Negele}},\ }\href {\doibase
  10.1103/PhysRevD.49.6039} {\bibfield  {journal} {\bibinfo  {journal} {Phys.
  Rev. D}\ }\textbf {\bibinfo {volume} {49}},\ \bibinfo {pages} {6039}
  (\bibinfo {year} {1994})},\ \Eprint {http://arxiv.org/abs/hep-lat/9312071}
  {arXiv:hep-lat/9312071} \BibitemShut {NoStop}%
\bibitem [{\citenamefont {Ji}\ \emph {et~al.}(2020)\citenamefont {Ji},
  \citenamefont {Liu}, \citenamefont {Liu}, \citenamefont {Zhang},\ and\
  \citenamefont {Zhao}}]{Ji:2020ect}%
  \BibitemOpen
  \bibfield  {author} {\bibinfo {author} {\bibfnamefont {X.}~\bibnamefont
  {Ji}}, \bibinfo {author} {\bibfnamefont {Y.-S.}\ \bibnamefont {Liu}},
  \bibinfo {author} {\bibfnamefont {Y.}~\bibnamefont {Liu}}, \bibinfo {author}
  {\bibfnamefont {J.-H.}\ \bibnamefont {Zhang}}, \ and\ \bibinfo {author}
  {\bibfnamefont {Y.}~\bibnamefont {Zhao}},\ }\href@noop {} {\  (\bibinfo
  {year} {2020})},\ \Eprint {http://arxiv.org/abs/2004.03543} {arXiv:2004.03543
  [hep-ph]} \BibitemShut {NoStop}%
\bibitem [{\citenamefont {Shuryak}\ and\ \citenamefont
  {Zahed}(2020)}]{Shuryak:2020ktq}%
  \BibitemOpen
  \bibfield  {author} {\bibinfo {author} {\bibfnamefont {E.}~\bibnamefont
  {Shuryak}}\ and\ \bibinfo {author} {\bibfnamefont {I.}~\bibnamefont
  {Zahed}},\ }\href@noop {} {\  (\bibinfo {year} {2020})},\ \Eprint
  {http://arxiv.org/abs/2008.06169} {arXiv:2008.06169 [hep-ph]} \BibitemShut
  {NoStop}%
\bibitem [{\citenamefont {Shuryak}(1982)}]{Shuryak:1981ff}%
  \BibitemOpen
  \bibfield  {author} {\bibinfo {author} {\bibfnamefont {E.~V.}\ \bibnamefont
  {Shuryak}},\ }\href {\doibase 10.1016/0550-3213(82)90478-3} {\bibfield
  {journal} {\bibinfo  {journal} {Nucl. Phys. B}\ }\textbf {\bibinfo {volume}
  {203}},\ \bibinfo {pages} {93} (\bibinfo {year} {1982})}\BibitemShut
  {NoStop}%
\bibitem [{\citenamefont {Sch\"afer}\ and\ \citenamefont
  {Shuryak}(1998)}]{Schafer:1996wv}%
  \BibitemOpen
  \bibfield  {author} {\bibinfo {author} {\bibfnamefont {T.}~\bibnamefont
  {Sch\"afer}}\ and\ \bibinfo {author} {\bibfnamefont {E.~V.}\ \bibnamefont
  {Shuryak}},\ }\href {\doibase 10.1103/RevModPhys.70.323} {\bibfield
  {journal} {\bibinfo  {journal} {Rev. Mod. Phys.}\ }\textbf {\bibinfo {volume}
  {70}},\ \bibinfo {pages} {323} (\bibinfo {year} {1998})},\ \Eprint
  {http://arxiv.org/abs/hep-ph/9610451} {arXiv:hep-ph/9610451} \BibitemShut
  {NoStop}%
\bibitem [{\citenamefont {Athenodorou}\ \emph {et~al.}(2018)\citenamefont
  {Athenodorou}, \citenamefont {Boucaud}, \citenamefont {De~Soto},
  \citenamefont {Rodr\'\i{}guez-Quintero},\ and\ \citenamefont
  {Zafeiropoulos}}]{Athenodorou:2018jwu}%
  \BibitemOpen
  \bibfield  {author} {\bibinfo {author} {\bibfnamefont {A.}~\bibnamefont
  {Athenodorou}}, \bibinfo {author} {\bibfnamefont {P.}~\bibnamefont
  {Boucaud}}, \bibinfo {author} {\bibfnamefont {F.}~\bibnamefont {De~Soto}},
  \bibinfo {author} {\bibfnamefont {J.}~\bibnamefont
  {Rodr\'\i{}guez-Quintero}}, \ and\ \bibinfo {author} {\bibfnamefont
  {S.}~\bibnamefont {Zafeiropoulos}},\ }\href {\doibase
  10.1007/JHEP02(2018)140} {\bibfield  {journal} {\bibinfo  {journal} {JHEP}\
  }\textbf {\bibinfo {volume} {02}},\ \bibinfo {pages} {140} (\bibinfo {year}
  {2018})},\ \Eprint {http://arxiv.org/abs/1801.10155} {arXiv:1801.10155
  [hep-lat]} \BibitemShut {NoStop}%
\bibitem [{\citenamefont {Hasenfratz}(2000)}]{Hasenfratz:1999ng}%
  \BibitemOpen
  \bibfield  {author} {\bibinfo {author} {\bibfnamefont {A.}~\bibnamefont
  {Hasenfratz}},\ }\href {\doibase 10.1016/S0370-2693(00)00105-2} {\bibfield
  {journal} {\bibinfo  {journal} {Phys. Lett. B}\ }\textbf {\bibinfo {volume}
  {476}},\ \bibinfo {pages} {188} (\bibinfo {year} {2000})},\ \Eprint
  {http://arxiv.org/abs/hep-lat/9912053} {arXiv:hep-lat/9912053} \BibitemShut
  {NoStop}%
\bibitem [{\citenamefont {Shuryak}(1999)}]{Shuryak:1999fe}%
  \BibitemOpen
  \bibfield  {author} {\bibinfo {author} {\bibfnamefont {E.~V.}\ \bibnamefont
  {Shuryak}},\ }\href@noop {} {\  (\bibinfo {year} {1999})},\ \Eprint
  {http://arxiv.org/abs/hep-ph/9909458} {arXiv:hep-ph/9909458} \BibitemShut
  {NoStop}%
\bibitem [{\citenamefont {Kock}\ \emph {et~al.}(2020)\citenamefont {Kock},
  \citenamefont {Liu},\ and\ \citenamefont {Zahed}}]{Kock:2020frx}%
  \BibitemOpen
  \bibfield  {author} {\bibinfo {author} {\bibfnamefont {A.}~\bibnamefont
  {Kock}}, \bibinfo {author} {\bibfnamefont {Y.}~\bibnamefont {Liu}}, \ and\
  \bibinfo {author} {\bibfnamefont {I.}~\bibnamefont {Zahed}},\ }\href
  {\doibase 10.1103/PhysRevD.102.014039} {\bibfield  {journal} {\bibinfo
  {journal} {Phys. Rev. D}\ }\textbf {\bibinfo {volume} {102}},\ \bibinfo
  {pages} {014039} (\bibinfo {year} {2020})},\ \Eprint
  {http://arxiv.org/abs/2004.01595} {arXiv:2004.01595 [hep-ph]} \BibitemShut
  {NoStop}%
\bibitem [{\citenamefont {Balitsky}\ \emph {et~al.}(1993)\citenamefont
  {Balitsky}, \citenamefont {Beneke},\ and\ \citenamefont
  {Braun}}]{Balitsky:1993ki}%
  \BibitemOpen
  \bibfield  {author} {\bibinfo {author} {\bibfnamefont {I.~I.}\ \bibnamefont
  {Balitsky}}, \bibinfo {author} {\bibfnamefont {M.}~\bibnamefont {Beneke}}, \
  and\ \bibinfo {author} {\bibfnamefont {V.~M.}\ \bibnamefont {Braun}},\ }\href
  {\doibase 10.1016/0370-2693(93)90142-5} {\bibfield  {journal} {\bibinfo
  {journal} {Phys. Lett. B}\ }\textbf {\bibinfo {volume} {318}},\ \bibinfo
  {pages} {371} (\bibinfo {year} {1993})},\ \Eprint
  {http://arxiv.org/abs/hep-ph/9309217} {arXiv:hep-ph/9309217} \BibitemShut
  {NoStop}%
\bibitem [{\citenamefont {Moch}\ \emph {et~al.}(1997)\citenamefont {Moch},
  \citenamefont {Ringwald},\ and\ \citenamefont {Schrempp}}]{Moch:1996bs}%
  \BibitemOpen
  \bibfield  {author} {\bibinfo {author} {\bibfnamefont {S.}~\bibnamefont
  {Moch}}, \bibinfo {author} {\bibfnamefont {A.}~\bibnamefont {Ringwald}}, \
  and\ \bibinfo {author} {\bibfnamefont {F.}~\bibnamefont {Schrempp}},\ }\href
  {\doibase 10.1016/S0550-3213(97)00592-0} {\bibfield  {journal} {\bibinfo
  {journal} {Nucl. Phys. B}\ }\textbf {\bibinfo {volume} {507}},\ \bibinfo
  {pages} {134} (\bibinfo {year} {1997})},\ \Eprint
  {http://arxiv.org/abs/hep-ph/9609445} {arXiv:hep-ph/9609445} \BibitemShut
  {NoStop}%
\bibitem [{\citenamefont {Brown}\ \emph {et~al.}(1978)\citenamefont {Brown},
  \citenamefont {Carlitz}, \citenamefont {Creamer},\ and\ \citenamefont
  {Lee}}]{Brown:1977eb}%
  \BibitemOpen
  \bibfield  {author} {\bibinfo {author} {\bibfnamefont {L.~S.}\ \bibnamefont
  {Brown}}, \bibinfo {author} {\bibfnamefont {R.~D.}\ \bibnamefont {Carlitz}},
  \bibinfo {author} {\bibfnamefont {D.~B.}\ \bibnamefont {Creamer}}, \ and\
  \bibinfo {author} {\bibfnamefont {C.-k.}\ \bibnamefont {Lee}},\ }\href
  {\doibase 10.1103/PhysRevD.17.1583} {\bibfield  {journal} {\bibinfo
  {journal} {Phys. Rev. D}\ }\textbf {\bibinfo {volume} {17}},\ \bibinfo
  {pages} {1583} (\bibinfo {year} {1978})}\BibitemShut {NoStop}%
\bibitem [{\citenamefont {Andrei}\ and\ \citenamefont
  {Gross}(1978)}]{Andrei:1978xg}%
  \BibitemOpen
  \bibfield  {author} {\bibinfo {author} {\bibfnamefont {N.}~\bibnamefont
  {Andrei}}\ and\ \bibinfo {author} {\bibfnamefont {D.~J.}\ \bibnamefont
  {Gross}},\ }\href {\doibase 10.1103/PhysRevD.18.468} {\bibfield  {journal}
  {\bibinfo  {journal} {Phys. Rev. D}\ }\textbf {\bibinfo {volume} {18}},\
  \bibinfo {pages} {468} (\bibinfo {year} {1978})}\BibitemShut {NoStop}%
\bibitem [{\citenamefont {Faccioli}\ and\ \citenamefont
  {Shuryak}(2001)}]{Faccioli:2001ug}%
  \BibitemOpen
  \bibfield  {author} {\bibinfo {author} {\bibfnamefont {P.}~\bibnamefont
  {Faccioli}}\ and\ \bibinfo {author} {\bibfnamefont {E.~V.}\ \bibnamefont
  {Shuryak}},\ }\href {\doibase 10.1103/PhysRevD.64.114020} {\bibfield
  {journal} {\bibinfo  {journal} {Phys. Rev. D}\ }\textbf {\bibinfo {volume}
  {64}},\ \bibinfo {pages} {114020} (\bibinfo {year} {2001})},\ \Eprint
  {http://arxiv.org/abs/hep-ph/0106019} {arXiv:hep-ph/0106019} \BibitemShut
  {NoStop}%
\bibitem [{\citenamefont {Dunne}\ and\ \citenamefont
  {\"Unsal}(2014)}]{Dunne:2013ada}%
  \BibitemOpen
  \bibfield  {author} {\bibinfo {author} {\bibfnamefont {G.~V.}\ \bibnamefont
  {Dunne}}\ and\ \bibinfo {author} {\bibfnamefont {M.}~\bibnamefont
  {\"Unsal}},\ }\href {\doibase 10.1103/PhysRevD.89.041701} {\bibfield
  {journal} {\bibinfo  {journal} {Phys. Rev. D}\ }\textbf {\bibinfo {volume}
  {89}},\ \bibinfo {pages} {041701} (\bibinfo {year} {2014})},\ \Eprint
  {http://arxiv.org/abs/1306.4405} {arXiv:1306.4405 [hep-th]} \BibitemShut
  {NoStop}%
\bibitem [{\citenamefont {Vandoren}\ and\ \citenamefont {van
  Nieuwenhuizen}(2008)}]{Vandoren:2008xg}%
  \BibitemOpen
  \bibfield  {author} {\bibinfo {author} {\bibfnamefont {S.}~\bibnamefont
  {Vandoren}}\ and\ \bibinfo {author} {\bibfnamefont {P.}~\bibnamefont {van
  Nieuwenhuizen}},\ }\href@noop {} {\  (\bibinfo {year} {2008})},\ \Eprint
  {http://arxiv.org/abs/0802.1862} {arXiv:0802.1862 [hep-th]} \BibitemShut
  {NoStop}%
\end{thebibliography}%

\end{document}